\documentclass{emulateapj}

\usepackage{graphicx}
\usepackage{epsfig}
\usepackage{epstopdf}
\usepackage{times}
\usepackage[]{natbib}
\usepackage{url}
\usepackage{color}
\usepackage{amssymb,amsmath}

\usepackage[varg]{txfonts}
\usepackage{color}           
\usepackage[table,xcdraw,dvipsnames]{xcolor}   
\usepackage[inline]{enumitem} 

\newcommand{\kms}{km~s$^{-1}\,$}

\shorttitle{Recurrent CME-like eruption - Part I}
\shortauthors{Syntelis et al.}

\begin{document}

\title{Recurrent CME-like eruptions in emerging flux regions. I. On the mechanism of eruptions.}

\author{P. Syntelis\altaffilmark{1} and V. Archontis\altaffilmark{1} and K. Tsinganos \altaffilmark{2}}
\affil{School of Mathematics and Statistics, St. Andrews University, St. Andrews, KY16 9SS, UK}
\affil{Section of Astrophysics, Astronomy and Mechanics, Department of Physics, University of Athens, Panepistimiopolis, Zografos 15784, Athens, Greece}

\email{ps84@st-andrews.ac.uk}


\begin{abstract}
We report on three-dimensional (3D) Magnetohydrodynamic (MHD) simulations of recurrent eruptions in emerging flux regions. We find that reconnection of sheared fieldlines, along the polarity inversion line of an emerging bipolar region, leads to the formation of a new magnetic structure, which adopts the shape of a magnetic flux rope during its rising motion. Initially, the flux rope undergoes a slow-rise phase and, eventually, it experiences a fast-rise phase and ejective eruption towards the outer solar atmosphere. In total, four eruptions occur during the evolution of the system. For the first eruption, our analysis indicates that the torus instability initiates the eruption and that tether-cutting reconnection of the fieldlines, which envelope the flux rope, triggers the rapid acceleration of the eruptive field. For the following eruptions, we conjecture that it is the interplay between tether-cutting reconnection and torus instability, which causes the onset of the various phases. We show the 3D shape of the erupting fields, focusing more on how magnetic fieldlines reconnect during the eruptions. 
We find that when the envelope fieldlines reconnect mainly with themselves, hot and dense plasma is transferred closer to the core of the erupting flux rope. The same area appears to be cooler and less dense when the envelope fieldlines reconnect with neighboring sheared fieldlines.
The plasma density and temperature distribution, together with the rising speeds, the energies and the size of the erupting fields indicate that they may account for small-scale (mini) Coronal Mass Ejections (CMEs).
\end{abstract}

\keywords{Sun: activity -- Sun: interior --
                Sun: Magnetic fields --Magnetohydrodynamics (MHD) --methods: numerical
               }

\section{Introduction}

The formation of Active Regions (ARs) is often associated with the emergence of magnetic flux (EMF) from the solar interior \citep[e.g.][]{Parker_1955}. Many explosive phenomena observed on the Sun, such as flaring events and CMEs, are associated with ARs. In fact, it has been observed that a single AR can produce several CMEs in a recurrent manner \citep[e.g. ][]{Nitta_etal2001, Zhang_etal2008,Wang_etal2013}. 

Solar eruptions have been studied extensively in the past. Observational studies have reported on the pre-eruptive phase of the eruption \citep[e.g.][]{Canou_Amari2010,Vourlidas_etal2012,Syntelis_etal2016}, the triggering of the eruptions \citep[e.g.][]{Zuccarello_etal2014,Reeves_etal2015,Chintzoglou_etal2015} and the propagation of the erupting structures in the interplanetary medium \citep[e.g.][]{Colaninno_etal2013} and towards the Earth \citep[e.g.][]{Patsourakos_etal2016}. 

Often, eruptions are associated with the formation of a twisted magnetic field structure, which is commonly referred to as a magnetic flux rope (FR) \citep[e.g.][]{Cheng_etal2011,Green_etal2011,Zhang_etal2012,Patsourakos_etal2013}. 
Still, various aspects regarding the process of formation, destabilization and eruption of FRs are up for debate.

Numerical models studying the formation of magnetic FRs in the solar atmosphere have extensively demonstrated the role of shearing, rotation and reconnection of fieldlines in the buildup of magnetic twist. As an example, magnetic flux emergence experiments \citep[e.g.][]{Magara_etal2001,Fan_2009,Archontis_Torok2008} have shown that shearing motions along a polarity inversion line (PIL), can lead to reconnection of sheared fieldlines and the gradual formation of FRs, which may erupt in a confined or ejective manner \citep[e.g. ][]{Archontis_etal2012}. 
Furthermore, experiments where rotational motions are imposed at the photospheric boundary \citep[symmetric and asymmetric driving of polarities, ][]{DeVore_etal2008,Aulanier_etal2010} have shown that the shearing motions can form a pre-eruptive FR and destabilize the system.

Once a FR is formed, it may erupt in an ejective manner towards outer space \citep[e.g. ][]{Leake_etal2014} or remained confined, for instance, by a strong overlying field \citep[e.g. ][]{Leake_etal2013}. There are two main proposed mechanisms, which might be responsible for the triggering and/or driving of the eruption of magnetic FRs. One is the non-ideal process of magnetic reconnection and the other is the action of an ideal MHD instability.

One example of reconnection which leads to the eruption of a magnetic FR, is the well-known tether-cutting mechanism \citep{Moore_etal1980,Moore_etal1992}. During this process, the footpoints of sheared fieldines reconnect along a PIL, forming a FR. The FR slowly rises dragging in magnetic field from the sides and a current sheet is formed underneath the FR. Eventually, fast reconnection of the fieldlines that envelope the FR occurs at the current sheet. Then, the upward reconnection outflow assists to the further rise of the FR. In this way, an imbalance is achieved between a) the upward magnetic pressure and tension force and b) the downward tension force of the envelope fieldlines. This leads to an ejective eruption of the FR.  Another example is the so-called break-out reconnection, between the envelope field and a pre-existing magnetic field. If the relative orientation of the two fields is antiparallel, (external) reconnection between them becomes very effective when they come into contact 
\citep[e.g. ][]{Antiochos_etal1999, Karpen_etal2012, Archontis_etal2012, Leake_etal2014}. 
This reconnection releases the downward magnetic tension of the envelope field and the FR can ``break-out'', experiencing an ejective eruption. We should highlight that the relative orientation and field strengths of the interacting magnetic systems are important parameters that affect the eruption of the FR. 
In previous studies, it has been shown that depending on the value of these parameters, the rising FR could experience an ejective eruption or be confined by the envelope field or even become annihilated by the interaction with the pre-existing magnetic field \citep[e.g. ][]{Galsgaard_etal2007, Archontis_etal2012, Leake_etal2014}.

Solar eruptions can also be triggered by ideal processes. For instance, the helical kink instability \citep[][]{Anzer_1968,Torok_etal2004}, which occurs when the twist of the FR exceeds a critical value that depends on the configuration of the FR (e.g. cylindrical, toroidal) and the line-tying effect \citep[e.g. ][]{Hood_Priest_1981,Torok_etal2004}. During the instability, the axis of the rising FR develops a helical shape. The eruption of the helical magnetic field could be ejective or confined, depending e.g. on how strong the overlying magnetic field is \citep{Torok_Kliem_2005}. 

Another crucial parameter, which affects the eruption of a FR is how the external constraining magnetic field drops along the direction of height. 
This is related to the so-called torus instability \citep{Bateman_1978,Kliem_etal2006} . In this model, a toroidal current ring with major radius $R$ is placed inside an external magnetic field. This external magnetic field drops along the direction of the major radius as $R^{-n}$. 
Due to the current ring's curvature, a hoop force acts on the current ring. This force is directed away from the center of the torus.
An inwards Lorentz force acts on the current ring due to the external magnetic field.
Previous studies (\citet{Bateman_1978,Kliem_etal2006}) showed that, if the decrease rate of the external field (i.e. $n=- \partial B_{external} / \partial \ln R$) exceeds a critical value ($n_{crit}=1.5$), the current ring becomes unstable. The decrease rate of the external field is commonly referred to as torus or decay index.

The range of values of the critical torus index is still under debate. For instance, studies of emerging flux tubes with an initial arch-like configuration, have reported higher values of the torus index \citep[$n=1.7-2$, ][]{Fan_etal2007,Fan_2010}. 
\citet{An_Magara_2013}, in a flux emergence simulation of a straight, horizontal flux tube, reported values of torus index well above 2.
\citet{Demoulin_etal2010} have found that the torus index can vary depending on a range of parameters, such as the thickness of the current channel (the axial current of a twisted FR is a current channel). In cases of thin current channels, the index was found to be 1 (1.5) for straight (circular) channels.  Also, the FR expansion during its eruption affects the critical value of torus instability. For thick channels, the critical index for circular and straight channels does not vary much. It takes values ranging from 1.1-1.3 (with expansion of the FR) and 1.2-1.5 (without expansion).  \citet{Zuccarello_etal2015} investigated the role of line-tying effects on the eruption. They performed a series of simulations with a setup similar to \citet{Aulanier_etal2010}, but with different velocity drivers at the photosphere. They found that the critical index did not depend greatly on the pre-eruptive photospheric motions, and it was found to take values within the range of 1.1-1.3.

In our paper, we show the results of a simulation of magnetic flux emergence, which occurs dynamically from the solar interior to the outer solar atmosphere. We focus on the formation of magnetic FRs in the emerging flux region and their possible eruption. In particular, we show how reconnection leads to the formation of the FRs and how / why these FRs erupt. We find that the emergence of a single sub-photospheric magnetic flux tube can drive recurrent eruptions, which are produced due to the combined action of the torus instability and reconnection of the envelope fieldlines in a tether-cutting manner. We find that, at least in the first eruption, the fast ejection phase of the torus unstable FR is triggered by tether-cutting reconnection.
A geometrical extrapolation of the size of the eruptions showed that they can develop into large-scale structures, with a size comparable to small CMEs. The plasma density and temperature distributions reveal that the structure of the erupting fields consist of three main parts: a ``core'', a ``cavity'' and a ``front edge'', which is reminiscent of the ``three-part'' structure of CMEs.
We find that the plasma, at the close vicinity of the ``core'', is hotter and denser when the envelope fieldlines reconnect with themselves in a tether-cutting manner during the eruption. The same area appears to be cooler and less dense, when the envelope fieldlines reconnect with some other neighboring (e.g. sheared J-like) fieldlines.

In Sec.~\ref{sec:initial_conditions} we describe the initial conditions of our simulations. Sec.~\ref{sec:overview} is an overview of the dynamics occurring in our simulation leading to four recurrent eruptions. In Sec.~\ref{sec:eruptions_mechanims} we show the morphology of the magnetic field (before, during and after the eruptions) and the triggering mechanism of these eruptions. In Sec.~\ref{sec:temp_etc} we show the distribution of various properties of the erupting fields, such as density, temperature, velocity and current profiles.
In Sec.~\ref{sec:extrapolation} we perform an extrapolation of the size of the erupting structures. In Sec.~\ref{sec:conclusions} we summarize the results.

\section{Numerical Setup}
\label{sec:initial_conditions}

\begin{figure}
    \centering
    \includegraphics[width=0.85\columnwidth]{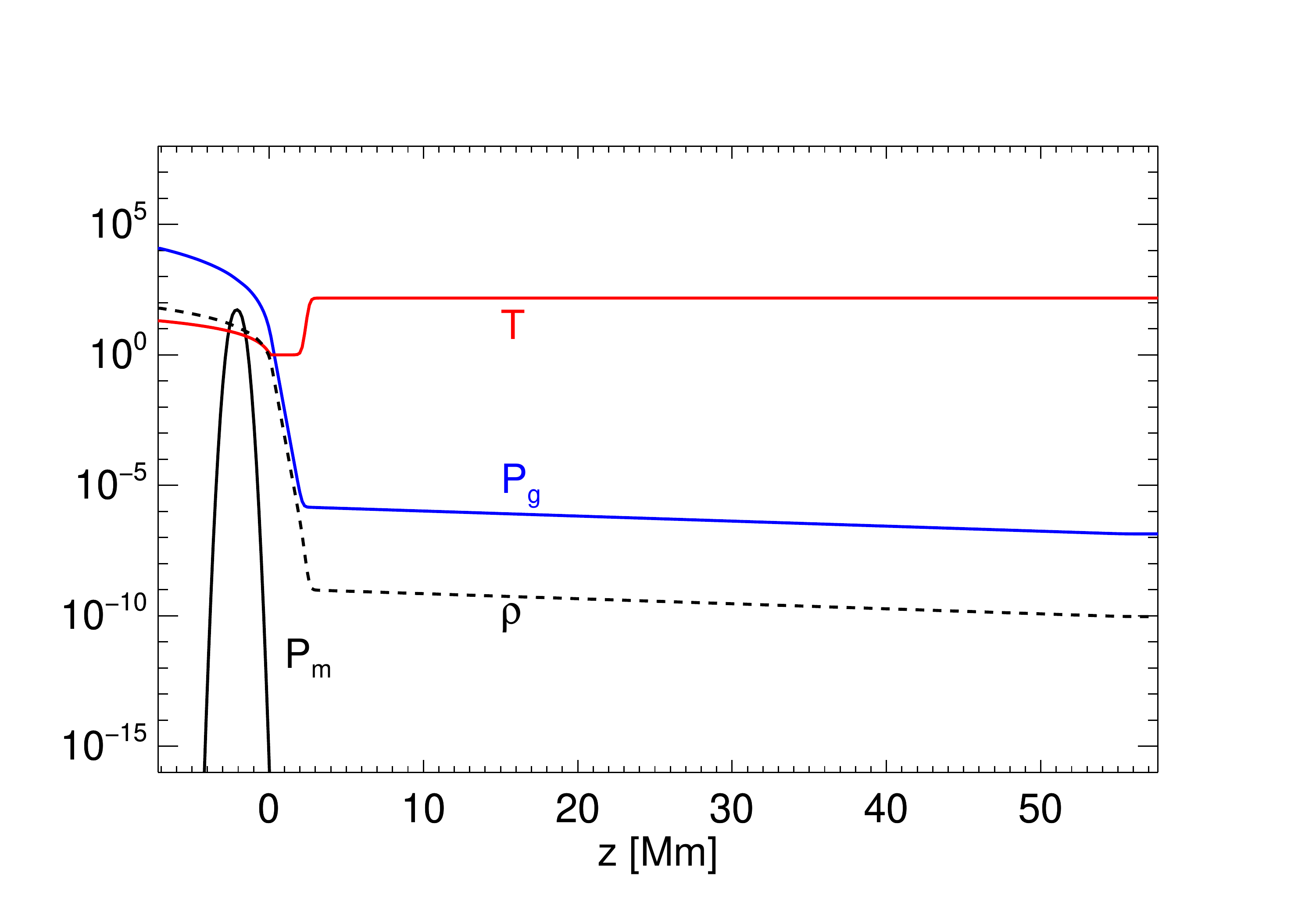}
    \caption{ Initial stratification of the background atmosphere in our simulation, in dimensionless units (temperature (T), density ($\rho$), magnetic pressure ($P_m$) and gas pressure ($P_g$)).
    }
    \label{fig:stratification}
\end{figure}

\begin{figure*}
\centering
\includegraphics[width=0.95\textwidth]{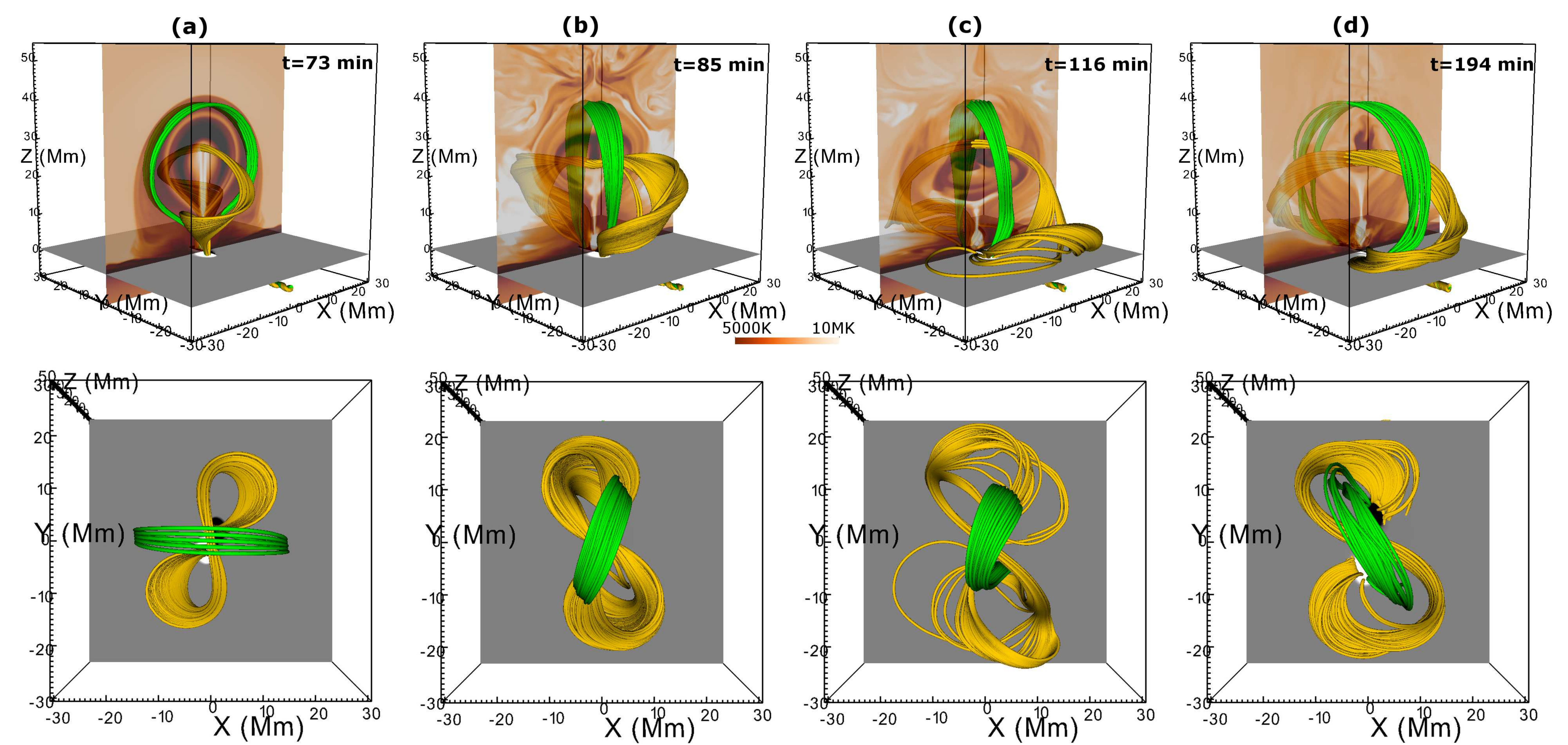}
\caption{Top: magnetic field line morphology and temperature distribution at the $xz$-midplane during the four eruptions of the simulation, at $t=73, 85, 116, 194$~min (for panels a, b, c and d respectively). Bottom: The same in a top-view. Two sets of fieldlines are shown: yellow (traced from the FR center) and green (traced from the envelope field). The horizontal $xy$-plane shows the distribution of $B_{z}$ at the photosphere (white:positive $B_{z}$, black:negative $B_{z}$, from -300~G to 300~G).}
\label{fig:4_plot}
\end{figure*}

To perform the simulations, we numerically solve the 3D time-dependent, resistive, compressible MHD equations in Cartesian geometry using the Lare3D code of \citet{Arber_etal2001}. The equations in dimensionless form are:
\begin{align}
&\frac{\partial \rho}{\partial t}+ \nabla \cdot (\rho \mathbf{v})  =0 ,\\
&\frac{\partial (\rho \mathbf{v})}{\partial t}  = - \nabla \cdot (\rho \mathbf{v v})  + (\nabla \times \mathbf{B}) \times \mathbf{B} - \nabla P + \rho \mathbf{g} + \nabla \cdot \mathbf{S} , \\
&\frac{ \partial ( \rho \epsilon )}{\partial t} = - \nabla \cdot (\rho \epsilon \mathbf{v}) -P \nabla \cdot \mathbf{v}+ Q_\mathrm{joule}+ Q_\mathrm{visc}, \\
&\frac{\partial \mathbf{B}}{\partial t} = \nabla \times (\mathbf{v}\times \mathbf{B})+ \eta \nabla^2 \mathbf{B},\\
&\epsilon  =\frac{P}{(\gamma -1)\rho},
\end{align}
where $\rho$, $\mathbf{v}$, $\mathbf{B}$ and P are density, velocity vector, magnetic field vector and gas pressure. Gravity is included. We assume a perfect gas with specific heat of $\gamma=5/3$. Viscous heating $Q_\mathrm{visc}$ and Joule dissipation $Q_\mathrm{joule}$ are also included.
We use explicit anomalous resistivity that increases linearly when the current density exceeds a critical value $J_c$:
\begin{equation}
  \eta=\begin{cases}
    \eta_{b}, & \text{if $\left|J\right|<J_{c}$}.\\
    \eta_{b}+\eta_{0}\left(\frac{\left|J\right|}{J_{c}}-1\right), & \text{if $\left|J\right|>J_{c}$}.
  \end{cases}
\end{equation}
, where $\eta_b=0.01$ is the background resistivity, $J_c=0.005$ is the critical current and $\eta_0=0.01$. 

We use normalization based on the photospheric values of density $\rho_\mathrm{c}=1.67 \times 10^{-7}\ \mathrm{g}\ \mathrm{cm}^{-3}$, length $H_\mathrm{c}=180 \ \mathrm{km}$
and magnetic field strength $B_\mathrm{c}=300 \ \mathrm{G}$. From these, we get pressure $P_\mathrm{c}=7.16\times 10^3\ \mathrm{erg}\ \mathrm{cm}^{-3}$, temperature $T_\mathrm{c}=5100~\mathrm{K}$, velocity $v_\mathrm{0}=2.1\ \mathrm{km} \ \mathrm{s}^{-1}$ and time $t_\mathrm{0}=85.7\ \mathrm{s}$.

The computational box has a size of $64.8\times64.8\times64.8 \ \mathrm{Mm}$ in the $x$, $y$, $z$ directions, in a $417\times417\times417$ grid. We assume periodic boundary conditions in the $y$ direction. Open boundary conditions are at the two $yz$-plane boundaries and at top of the numerical box.
The domain consists of an adiabatically stratified sub-photosheric layer at $-7.2\ \mathrm{Mm}\le z < 0 \ \mathrm{Mm}$, an isothermal photospheric-chromospheric layer at $0 \ \mathrm{Mm} \le z < 1.8 \ \mathrm{Mm} $, a transition region  at $1.8 \ \mathrm{Mm} \le z < 3.2 \ \mathrm{Mm}$ and an isothermal coronal at $3.2 \ \mathrm{Mm} \le z < 57.6 \ \mathrm{Mm}$.
We assume to have a field-free atmosphere in hydrostatic equilibrium. The initial distribution of temperature (T), density ($\rho$), gas  ($P_\mathrm{g}$) pressure is shown in Fig. \ref{fig:stratification}.

We place a straight, horizontal FR at $z=-2.1 \ \mathrm{Mm}$. The axis of the FR is oriented along the $y$-direction, so the transverse direction is along $x$ and height is in the $z$-direction.
The magnetic field of the FR is:
\begin{align}
B_{y} &=B_\mathrm{0} \exp(-r^2/R^2), \\
B_{\phi} &= \alpha r B_{y}
\end{align}
where $R=450$~km a measure of the FR's radius, $r$ the radial distance from the FR's axis and $\alpha= 0.4$ ($0.0023$~km$^{-1}$) is a measure of twist per unit of length. 
The magnetic field's strength is $B_0=3150$~G.
Its magnetic pressure ($P_m$) is over-plotted in Fig.~\ref{fig:stratification}.
Initially the FR is in pressure equilibrium. The FR is destabilized by imposing a density deficit along it's axis, similar to the work by \citet{Archontis_etal2004}:
\begin{equation}
\Delta \rho = \frac{p_\mathrm{t}(r)}{p(z)} \rho(z) \exp(-y^2/\lambda^2),
\label{eq:deficit}
\end{equation}
where $p$ is the external pressure and  $p_\mathrm{t}$ is the total pressure within the FR. The parameter $\lambda$ is the length scale of the buoyant part of the FR. We use $\lambda=5$ ($0.9$~Mm).

\section{Recurrent Eruptions}

\subsection{Overall evolution: a brief overview}
\label{sec:overview}

In the following, we briefly describe the overall evolution of the emerging flux region during the running time of the simulation. At $t$=25~min, the crest of the sub-photospheric FR reaches the photosphere. It takes 10~min for the magnetic buoyancy instability criterion \citep[see][]{Acheson1979, Archontis_etal2004} to be satisfied and, and thus, for the first magnetic flux elements to emerge at and above the solar surface. Eventually, the emerging magnetized plasma expands as it rises, due to the magnetic pressure inside the tube and the decreasing gas pressure of the background stratified atmosphere. Because of the expansion, the outermost expanding fieldlines adopt a fan-like configuration, forming an envelope field that surrounds all the upcoming magnetized plasma. As we discuss later in this paper, the characteristics and dynamical evolution of this envelope field play an important role towards understanding the eruptions coming from the emerging flux region. 

At the photosphere, the emergence of the field forms a bipolar region with a strong PIL. Similarly to previous studies \citep[e.g.][]{Manchester_2001,Archontis_Torok2008, Leake_etal2013}, we find that the combined action of shearing, driven by the Lorentz force along the PIL, and reconnection of the sheared fieldlines, leads to the formation of a new magnetic FR, which eventually erupts towards the outer space. In fact, this is an ongoing process, which leads to the formation and eruption of several FRs during the evolution of the system. Since these FRs are formed after the initial flux emergence at the photosphere, we will refer to them as the post-emergence FRs. 

Fig.~\ref{fig:4_plot} shows the temperature distribution (vertical $xz$-midplane) and selected fieldlines at the times of four successive eruptions in our simulation (panels a-d). The temperature distribution delineates the (bubble-shaped) volume of the erupting field, which is filled by cool and hot plasma. In Sec.~\ref{sec:temp_etc}, we discuss the physical properties (e.g. temperature, density) of the erupting plasma in more detail. The fieldlines are drawn in order to show a first view of the shape of the envelope field (green) and the core of the erupting FRs (yellow). Notice the strongly azimuthal nature of the envelope field and the S-shaped configuration of the FR's fieldlines in the first eruption (Fig.~\ref{fig:4_plot}a, top view). 
In the following eruptions, the orientation of the envelope field changes (in a counter-clockwise manner, Fig.~\ref{fig:4_plot}b-d, top view).
The morphology of the fieldlines during the four eruptions is discussed in detail, in Sec.~\ref{sec:eruptions_mechanims}. We find that all the eruptions are fully ejective (i.e. they exit the numerical domain from the top boundary).

\begin{figure}
\centering
\includegraphics[width=0.95\columnwidth]{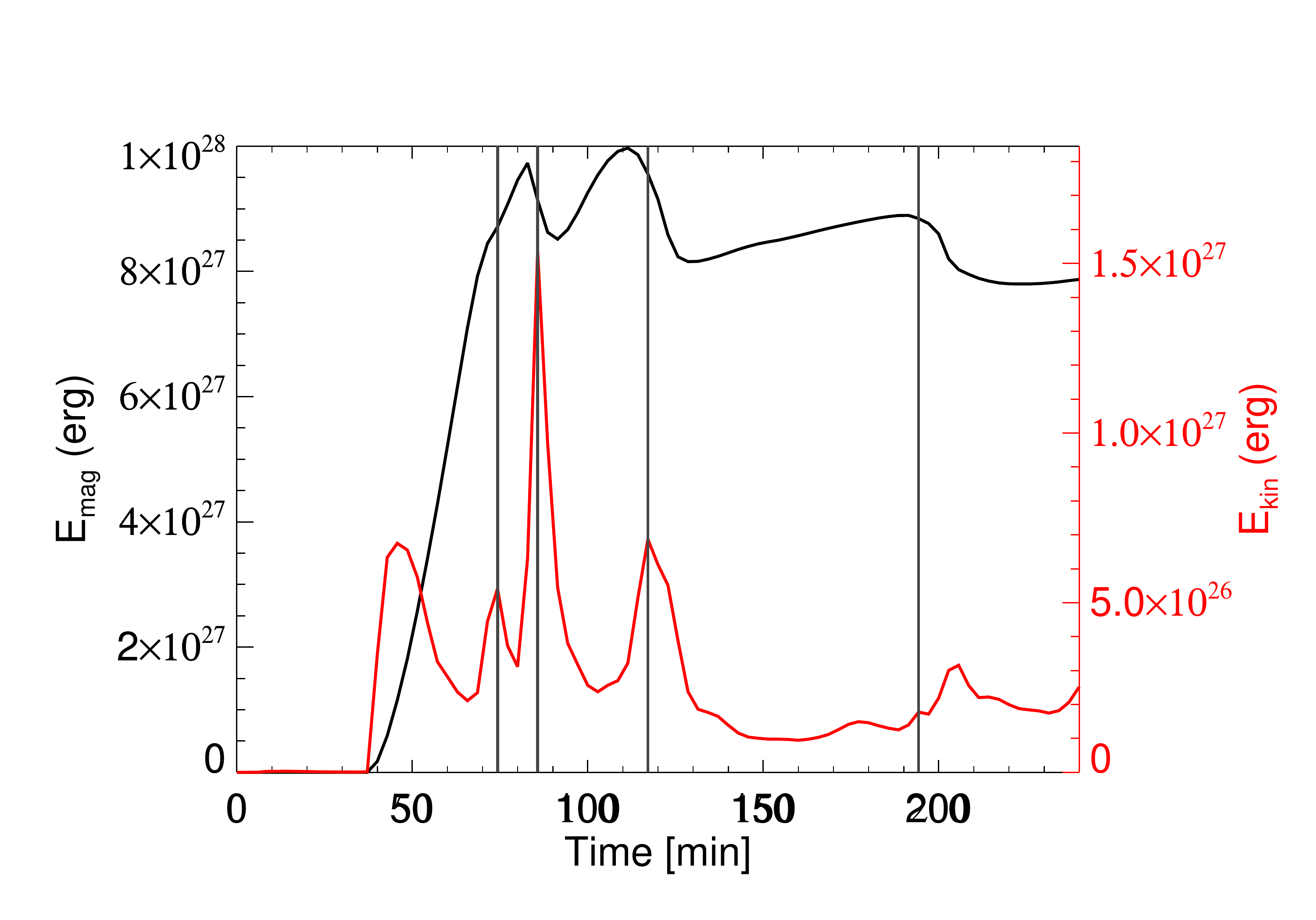}
\caption{ Magnetic (black) and kinetic (red) energy above the middle of the photosheric-chromospheric layer ($z=$1.37~Mm).  Vertical black lines mark the kinetic energy maxima related to the four eruptions.}
\label{fig:energy}
\end{figure}

To further describe the overall dynamical evolution of the eruptions, we calculate the total magnetic and kinetic energy (black and red line respectively, Fig.~\ref{fig:energy}) above the mid-photosphere ($z=1.37$~Mm). The first maximum of kinetic energy at $t=45.7$~min corresponds to the initial emergence of  the field. Then, we find four local maxima of the magnetic and kinetic energies, which correspond to the four eruptions (e.g. kinetic energy peaks at $t=74.3,\, 85.7,\, 117.1,\, 194.3$~min, marked by vertical lines in the figure). As expected, the magnetic (kinetic) energy decreases (increases) after each eruption. Notice that this is less pronounced for the magnetic energy in the first eruption because of the continuous emergence of magnetic flux, which increases the total amount of magnetic energy above the mid-photosphere. Also, the local maximum of the kinetic energy at $t=205.7$~min corresponds mainly to the fast reconnection upflow underneath the erupting FR, which is about to exit the numerical domain.

\begin{figure}
\centering
\includegraphics[width=0.95\columnwidth]{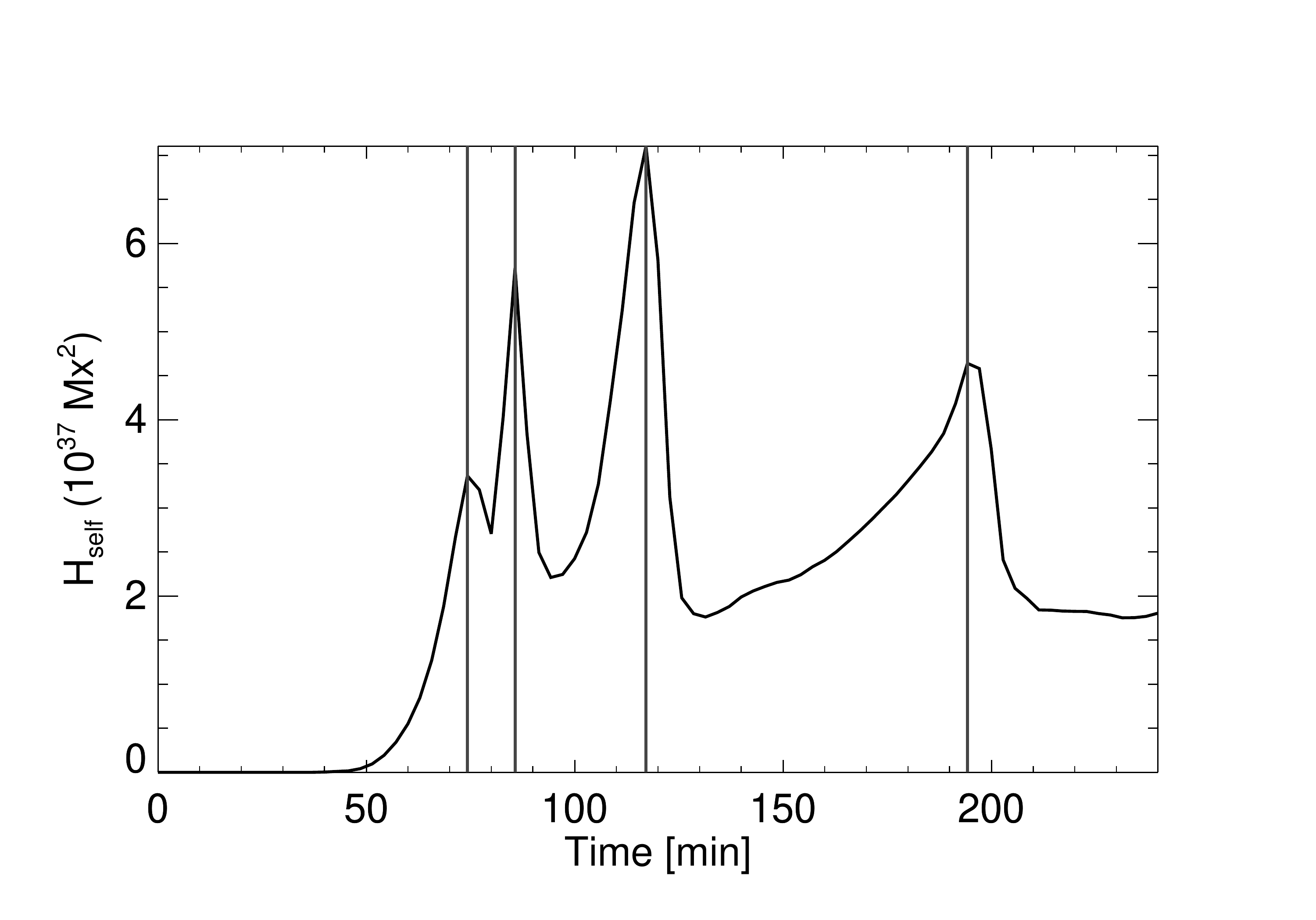}
\caption{Self helicity above the middle of the photosheric-chromospheric layer ($z=$1.37~Mm).  Vertical black lines mark the kinetic energy maxima related to the four eruptions.}
\label{fig:helicity}
\end{figure}

In a similar way, we compute the self helicity (Fig.~\ref{fig:helicity}). For a single twisted flux tube, the self-helicity is assumed to correspond to the twist within the flux tube.
For the calculation we used the method described in \citet{Moraitis_etal2014}. Overall, we find that the temporal evolution of the self-helicity is similar to that of the kinetic energy (e.g. they reach local maxima at the same time), which indicates that the erupted field is twisted. We also find that between the eruptions, self helicity increases because of the gradual build up of the twist of the post-emergence FRs.


\begin{figure*}
\centering
\includegraphics[width=0.70\textwidth]{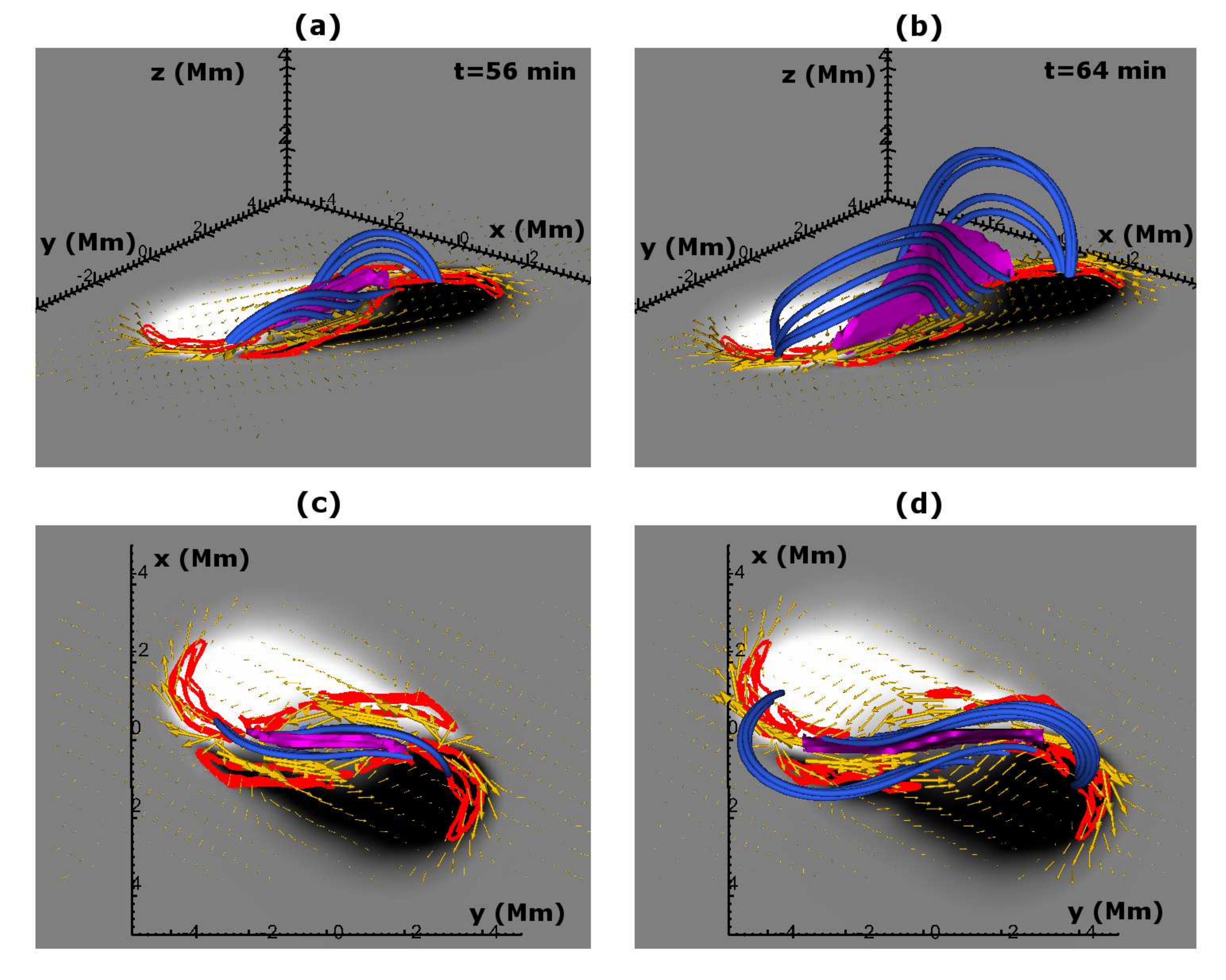}
\caption{Side and top views of the shape of selected fieldlines at $t$=56~min (a,c) and $t$=64~min (b,d). The horizontal slice shows the distribution of $B_{z}$ (in black and white, from -300~G to 300~G) at $z=0.7$~Mm. Yellow arrows represent the photospheric velocity field scaled by magnitude. Photospheric vorticity is shown by the red contours.  Purple isosurface shows $\left| J/B \right|>0.3$.}
\label{fig:jplot}
\end{figure*}

\subsection{Flux rope formation and eruption mechanisms}
\label{sec:eruptions_mechanims}
\subsubsection{First eruption}
\label{sec:eruption1}

\begin{figure*}
\centering
\includegraphics[width=0.75\textwidth]{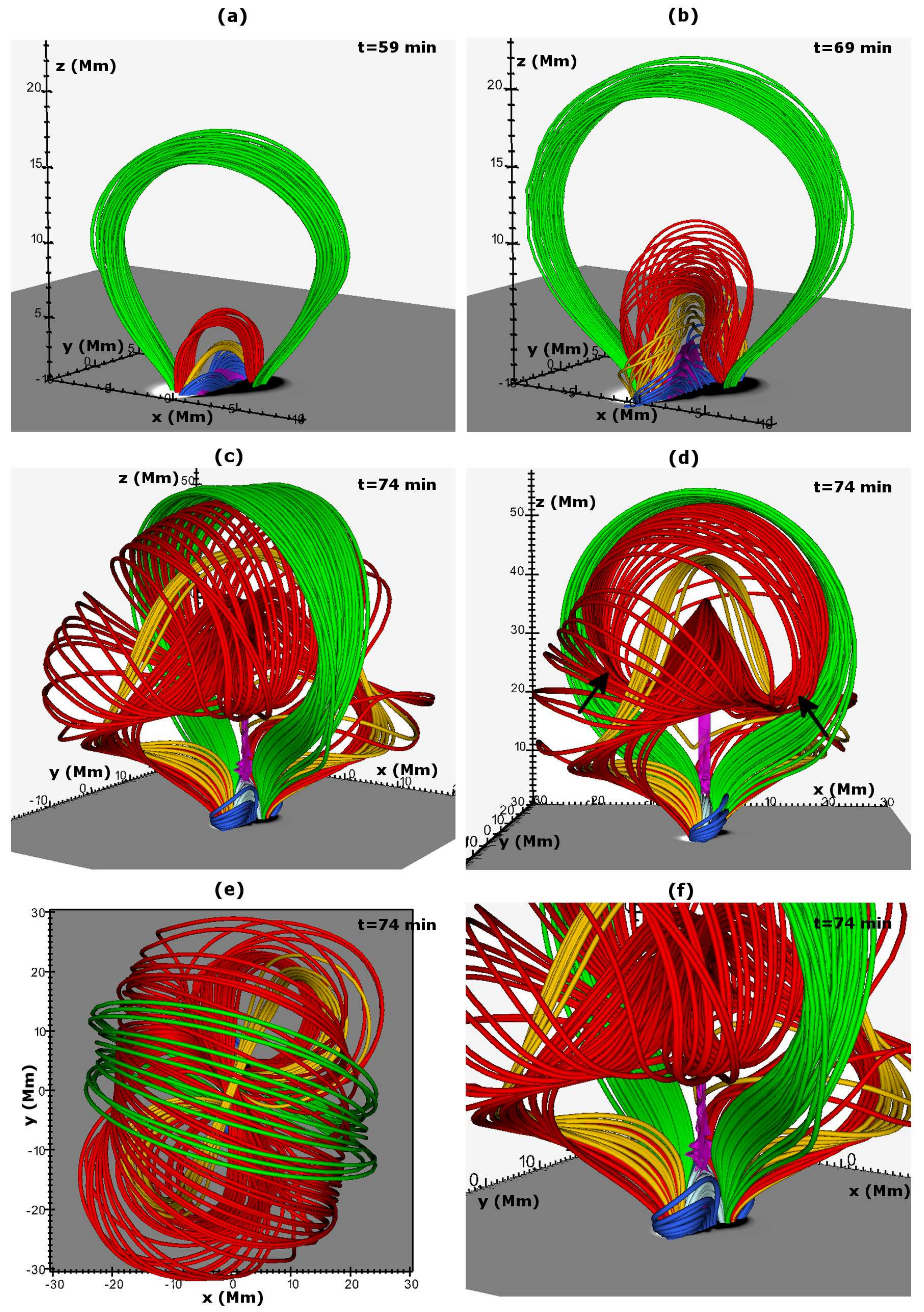}
\caption{Field line morphology of the first eruption at $t$= 59 (a), 69 (b), 74 (c-f)~min.  Green lines are traced from the top of the envelope field. Red lines are envelope fieldlines traced above the FR (c,d,e). Blue lines are J-shaped lines. Yellow lines are traced from the FR center.  Purple isosurface is $\left| J/B \right|>0.3$. \textbf{(c-e)}: $t$=74~min eruption from side, front and top view. \textbf{(d)}: Arrows show the two concave-upwards segments of the W-like (red) fieldlines. \textbf{(f)}: Close up of (c). Cyan lines illustrate the post-reconnection arcade.}
\label{fig:eruption1}
\end{figure*}

\begin{figure}
\centering
\includegraphics[width=0.90\columnwidth]{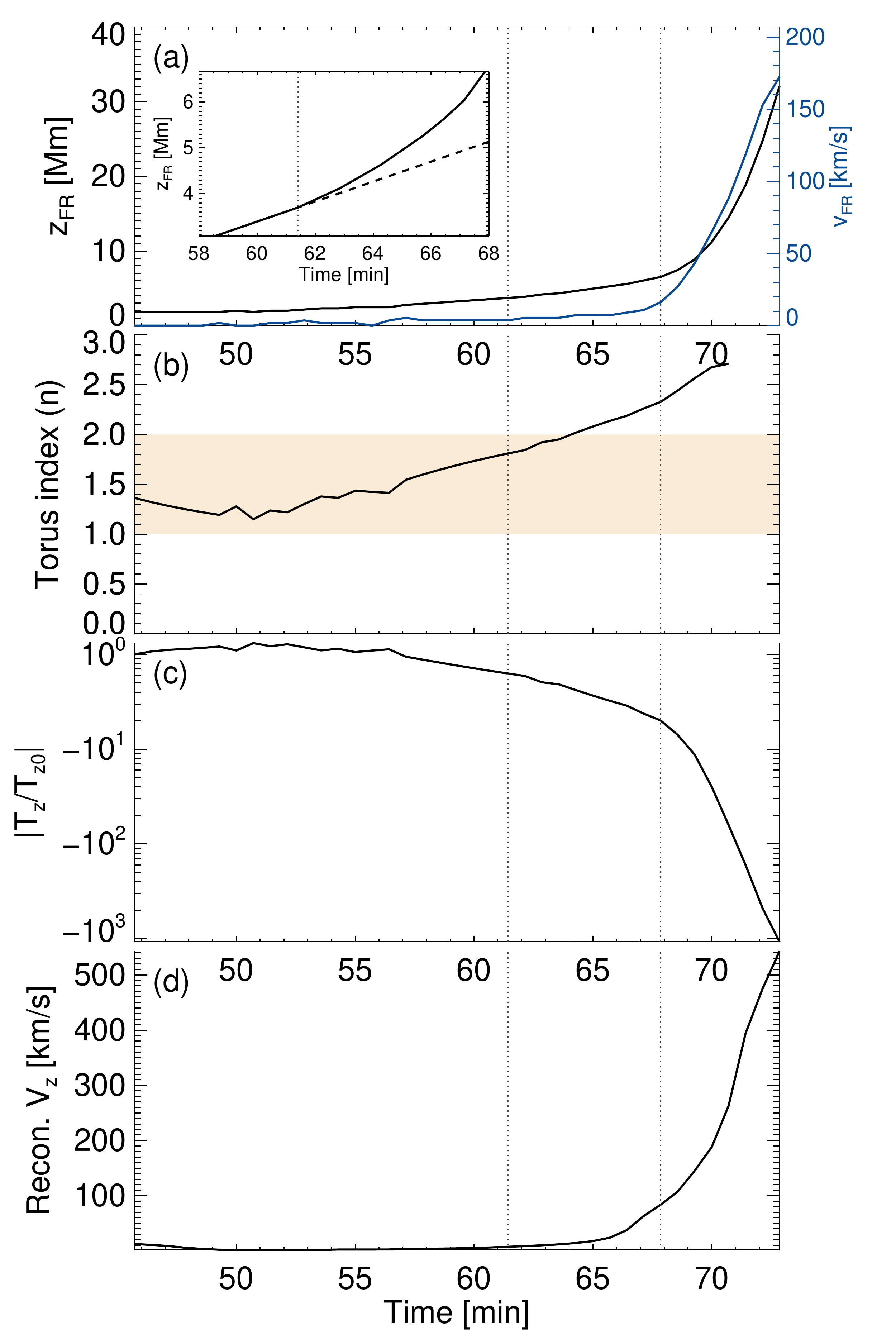}
\caption{First eruption's key parameters with time. 
\textbf{(a):} Height-time profile of FR center (black) and FR velocity (h-t derivative, blue). The insert shows a close-up of the height time profile for $t=58-68$~min.
\textbf{(b):}  Torus index measured at the FR center. The highlighted region shows an estimated range of values for the occurrence of a torus instability.
\textbf{(c):} Ratio of mean tension ($T_z$) over its initial value ($T_{z_0}$). Mean tension is measured from the apex of the FR to the top of the envelope field.
\textbf{(d):} Maximum $V_z$ of the reconnection outflow. Vertical lines mark the times of the possible onset of the torus instability (first line) and the tether-cutting reconnection of the envelope field (second line)}
\label{fig:vars_eruption1}
\end{figure}

The formation of the post-emergence FR occurs in the low atmosphere due to the combination of: a) shearing and converging motions along the PIL, b) rotation of the polarities of the emerging flux region and c) reconnection of the sheared and rotated fieldlines along the PIL. 

Firstly, we would like to focus on the role of shearing along the PIL and the rotation of the polarities during the pre-eruptive phase. For this reason, we present a side view (Fig.~\ref{fig:jplot}a,b) and a top view (Fig.~\ref{fig:jplot}c,d) of a close-up of the emerging flux region. We plot the sheared arcade fieldlines (blue), the $\left| J/B \right|$ isosurface and the photospheric $B_z$ component of the magnetic field (black/white plane). On the photospheric plane, we also plot the planar component of the velocity field vector (yellow arrows) and the $\omega_z$ component of vorticity (red contours). 
The visualization of the velocity field reveals: a) the shearing motion along the PIL (the yellow arrows are almost antiparallel on the two sides of the PIL) and b) the converging motions towards the PIL and close to the two main polarities, due to their rotation.
These motions (shear and rotation) are also apparent by looking at the vertical component of the vorticity (red contours).  Notice that $\omega_z$ is strong close to the two polarities, where the rotation is fast. Along and sideways of the PIL, there is only apparent ``vorticity'' due to the velocity, which is developed by the shearing.

The footpoints of the sheared arcade fieldlines are rooted at both sides of the PIL (e.g. blue lines in Fig.~\ref{fig:jplot}a,c). Due to the shearing, their footpoints move towards the two polarities where they undergo rotation (e.g. see the footpoints of the blue fieldlines, which go through the red contours close to the two opposite polarities, Fig.~\ref{fig:jplot}b,d).
Due to rotation, the sheared fieldlines  adopt the characteristic hook-shaped edge, forming J-like loops. The isosurface of high values of $\left| J/B\right|$ shows the formation of a strong current between the J-like loops. When the J-like fieldlines reconnect at the current sheet, new twisted fieldlines are formed, with an overall sigmoidal shape.

Figure~\ref{fig:eruption1} is a visualization of a series of selected fieldlines during the slow-rise (panels a and b) and the fast-rise (panels c-f) phase of the first eruption. In a similar manner to Fig.~\ref{fig:jplot}, Fig.~\ref{fig:eruption1}a shows the sheared fieldlines (blue) and the $\left| J/B \right|$ isosurface (purple). 
Reconnection between the sheared fieldlines forms a new set of longer fieldlines (yellow), which connect the distant footpoints of the sheared fieldlines. Thus, the longer fieldlines produce a magnetic loop above the PIL. As time goes on (panel b), further reconnection between the J-like sheared fieldlines (blue) form another set of fieldlines, which wrap around the magnetic loop, producing the first (post-emergence) magnetic FR. The red and green fieldlines are not reconnected fieldlines. They have been traced from arbitrary heights above the yellow fieldlines. They belong to the emerging field, which has expanded into the corona. In that respect, they create an envelope field for the new magnetic FR.

Eventually, the envelope fieldlines just above the FR  (e.g. red lines, Fig.~\ref{fig:eruption1}b) are stretched vertically and their lower segments come into contact and reconnect at the flare current sheet underneath the FR in a tether-cutting manner. Hereafter, for simplicity, we call the reconnection between envelope fieldlines as EE-TC reconnection (i.e. Envelope Envelope - Tether Cutting reconnection). This reconnection occurs in a fast manner, triggering an explosive acceleration of the FR. During this process, the plasma temperature at the flare current sheet reaches values up to 6~MK. The rapid eruption is followed by a similar type of reconnection of the outermost fieldlines of the envelope field (green lines, Figs.~\ref{fig:eruption1}c). Fig.~\ref{fig:eruption1}c,d,e show the side, front and top view of the fieldline morphology at $t$=74~min. Fig.~\ref{fig:eruption1}f is a close up of the reconnection site underneath the erupting FR.

Notice that, due to EE-TC reconnection, the red fieldlines are wrapped around the central region of the erupting field (yellow fieldlines). They make at least two turns around the axis, becoming part of the erupting FR. During the eruption, these fieldlines may reconnect more than once, and thus, have more than two full turns around the axis.
The close-up in Fig.~\ref{fig:eruption1}f shows that a post-reconnection arcade (light blue fieldlines) is formed below the flare current sheet. At the top of the arcade, the plasma is compressed and the temperature increases up to 10~MK.

The time evolution of the post-emergence FR can be followed by locating its axis at different times. To find the axis, we use a vertical 2d cut (at the middle of the FR, along its length), which is perpendicular to the fieldlines of the FR. 
Then we locate the maximum of the normal component of the magnetic field ($B_n$) on this 2d plane for every snapshot. 
We have also found that the location of the axis of the FR is almost identical to the location of maximum plasma density within the central region of the FR. 
The latter can be used as an alternative tracking-method for the location of the FR's axis. 
Using the above method(s), we are able to plot the height-time profile of the erupting FR (see Fig.~\ref{fig:vars_eruption1}a black line) and its derivative (blue line).
The h-t profile shows a phase of gradual upward motion (slow-rise phase), which is followed by an exponential period (fast-rise phase). The terminal velocity before the FR exits the numerical box is 170~\kms. During the eruptive phase, the FR is not very highly twisted and also it does not have the characteristic deformation of its axis that results from the kink instability. As a result, kink instability does not seem to play a role in this case. To study whether torus instability is at work, we follow the torus index calculation method of \citet{Fan_etal2007, Aulanier_etal2010}. 
We first estimate the external (envelope) field by calculating the potential magnetic field ($B_p$). 
This is done based on the calculations made to derive the helicity \citep[details in][]{Moraitis_etal2014}.
To solve the Laplace equation for the calculation of the potential field, it is assumed that both the magnetic field and the potential field have the same normal component at the boundaries (Neumann conditions). The lower $xy$-plane boundary is the photosphere at $z=0.51$~Mm and the rest of the boundaries are the sides of the numerical domain. Having calculated $B_p$, we then compute the torus index as $n=-z \partial \ln B_p / \partial z$. 
Then, we find the value of the torus index at the position of the FR center by measuring the value of $n$ along the h-t profile. We plot the results in Fig.~\ref{fig:vars_eruption1}b.

According to the height-time profile (black line and inset in Fig.~\ref{fig:vars_eruption1}a), we find that the FR enters an exponential rise phase just after $t=61.4$~min (first vertical line). The torus index at this time is $n=1.81$, which lies within the estimated range of values for the occurrence of the torus instability (see Introduction and the highlighted region in Fig.~\ref{fig:vars_eruption1}e). Therefore, we anticipate that the FR in our simulation becomes torus unstable at $t\geq 61.4$~min.

We should highlight that the envelope fieldlines above the FR start to reconnect in a TC manner at $t\geq 67.9$~min (second vertical line, Fig.~\ref{fig:vars_eruption1}). As a result, the mean tension of the envelope fieldlines (Fig.~\ref{fig:vars_eruption1}c) decreases while the FR height and velocity increase dramatically (Fig.~\ref{fig:vars_eruption1}a). We also find that the fast reconnection jet ($V_z$ up to $550$~\kms), which is ejected upward from the flare current sheet, is transferring momentum to the FR and contributes to its acceleration (Fig.~\ref{fig:vars_eruption1}d).

\begin{figure*}
\centering
\includegraphics[width=0.85\textwidth]{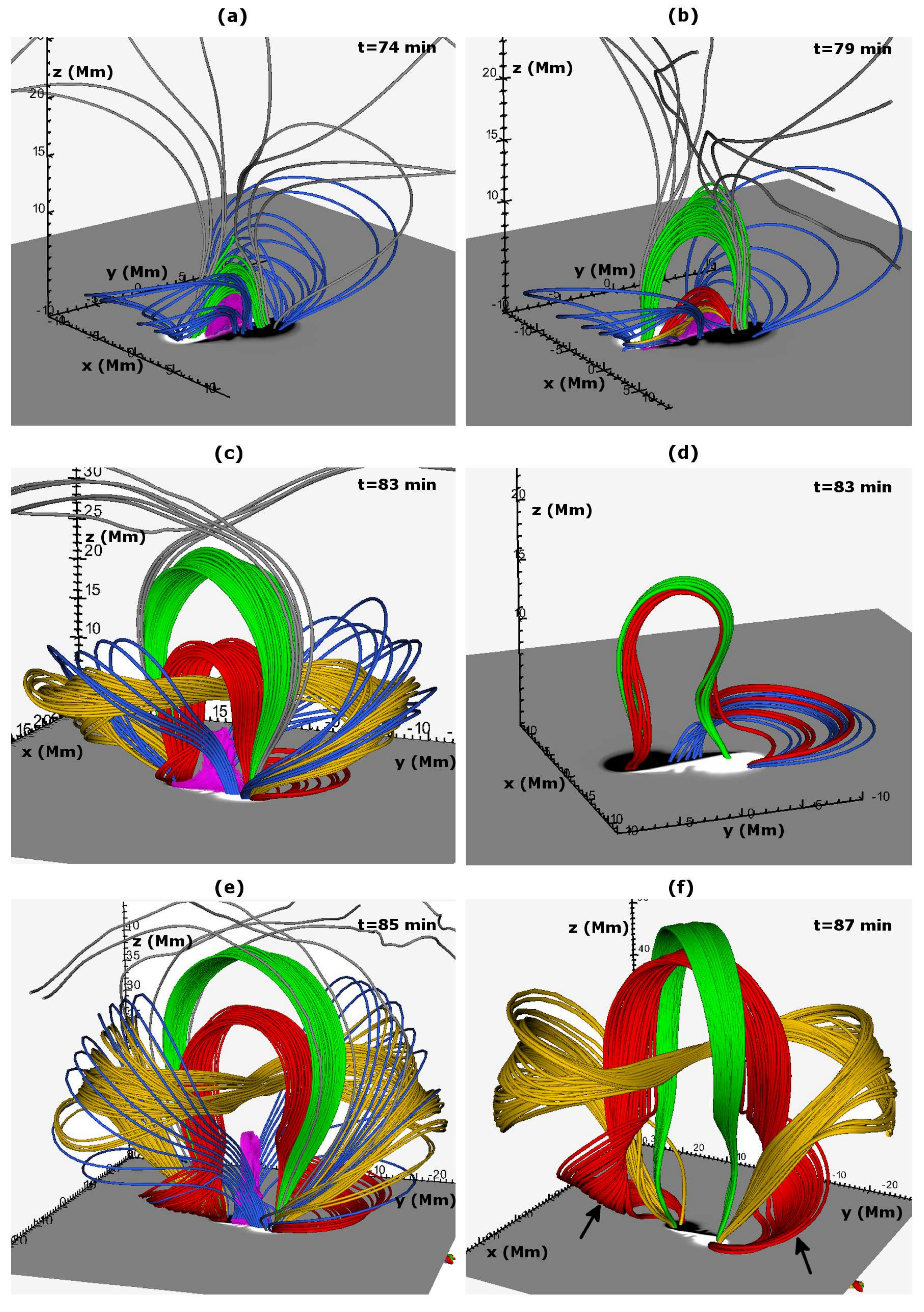}
\caption{ Field line morphology of the second eruption at $t$=74, 79, 83, 85, 87~min. Green lines are traced from the top of the post-reconnection arcade field of the first eruption. Red lines are envelope fieldlines traced above the FR (b,c,d,e,f). Blue lines are J-shaped lines. Yellow lines are traced from the FR center.  Purple isosurface is $\left| J/B \right|>0.3$. Gray lines are fieldlines from the first eruption (now acting as external field).
\textbf{(d):} Closeup of (c) showing the EJ-TC reconnection. \textbf{(f)}: Arrows show the two hook-shaped segments of the fieldlines (red lines). }
\label{fig:eruption2}
\end{figure*}
\begin{figure}
\centering
\includegraphics[width=0.90\columnwidth]{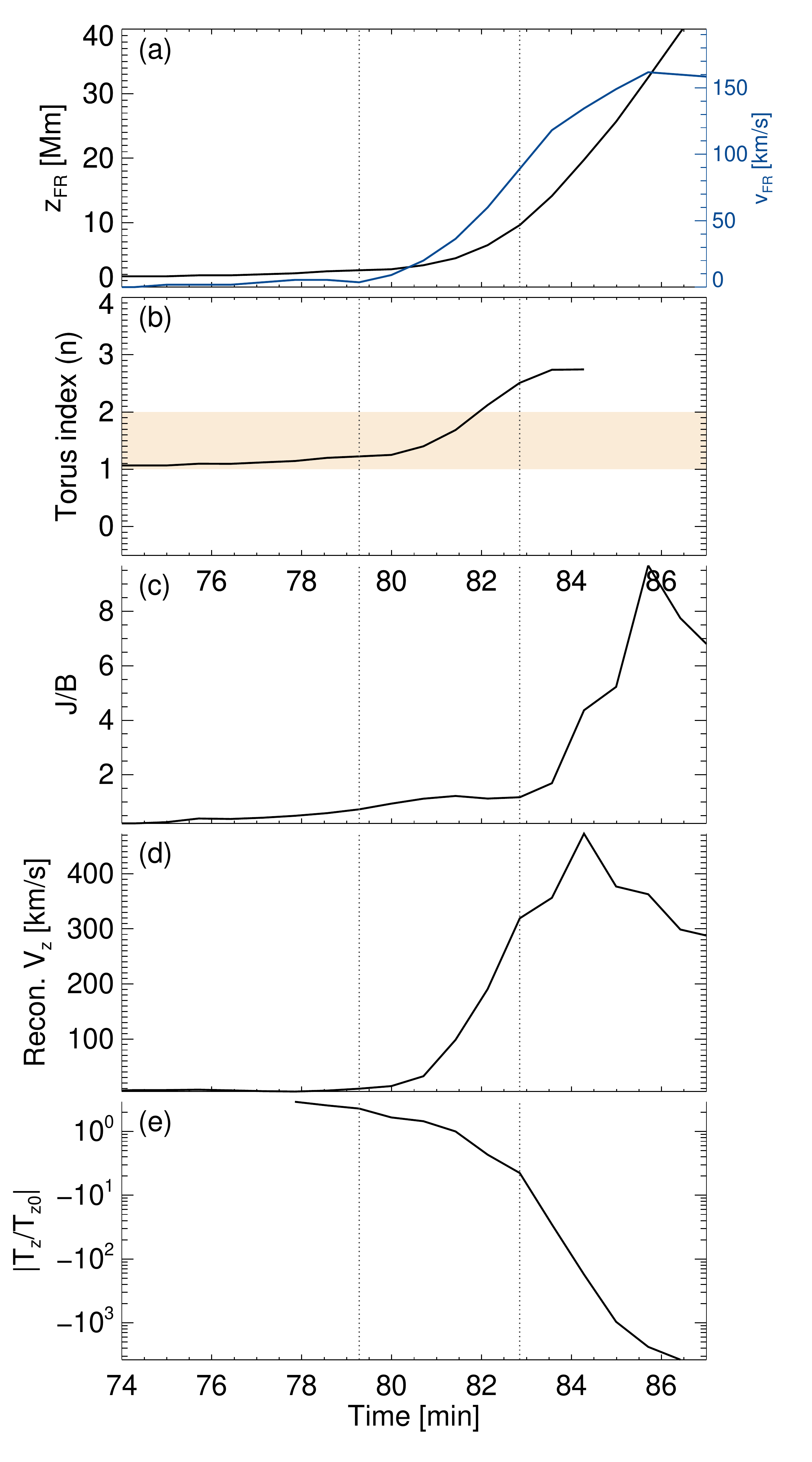}
\caption{Second eruption's key parameters with time.
\textbf{(a):} Height-time profile of FR center (black) and  FR velocity (h-t derivative, blue) \textbf{(b):} Torus index measured along the height-time profile. The highlighted region shows an estimated range of values for the occurrence of a torus instability. \textbf{(c):} Maximum $\left| J/B \right|$ along the CS. \textbf{(d):} Maximum reconnection outflow. \textbf{(e):} Ratio of mean tension ($T_z$) over its initial value ($T_{z_0}$). 
Vertical lines mark the times of the possible onset of the torus instability (first line) and the EJ-TC reconnection (second line).
}
\label{fig:vars_eruption2}
\end{figure}


\subsubsection{Second eruption}
\label{sec:eruption2}

\begin{figure*}
\centering
\includegraphics[width=0.85\textwidth]{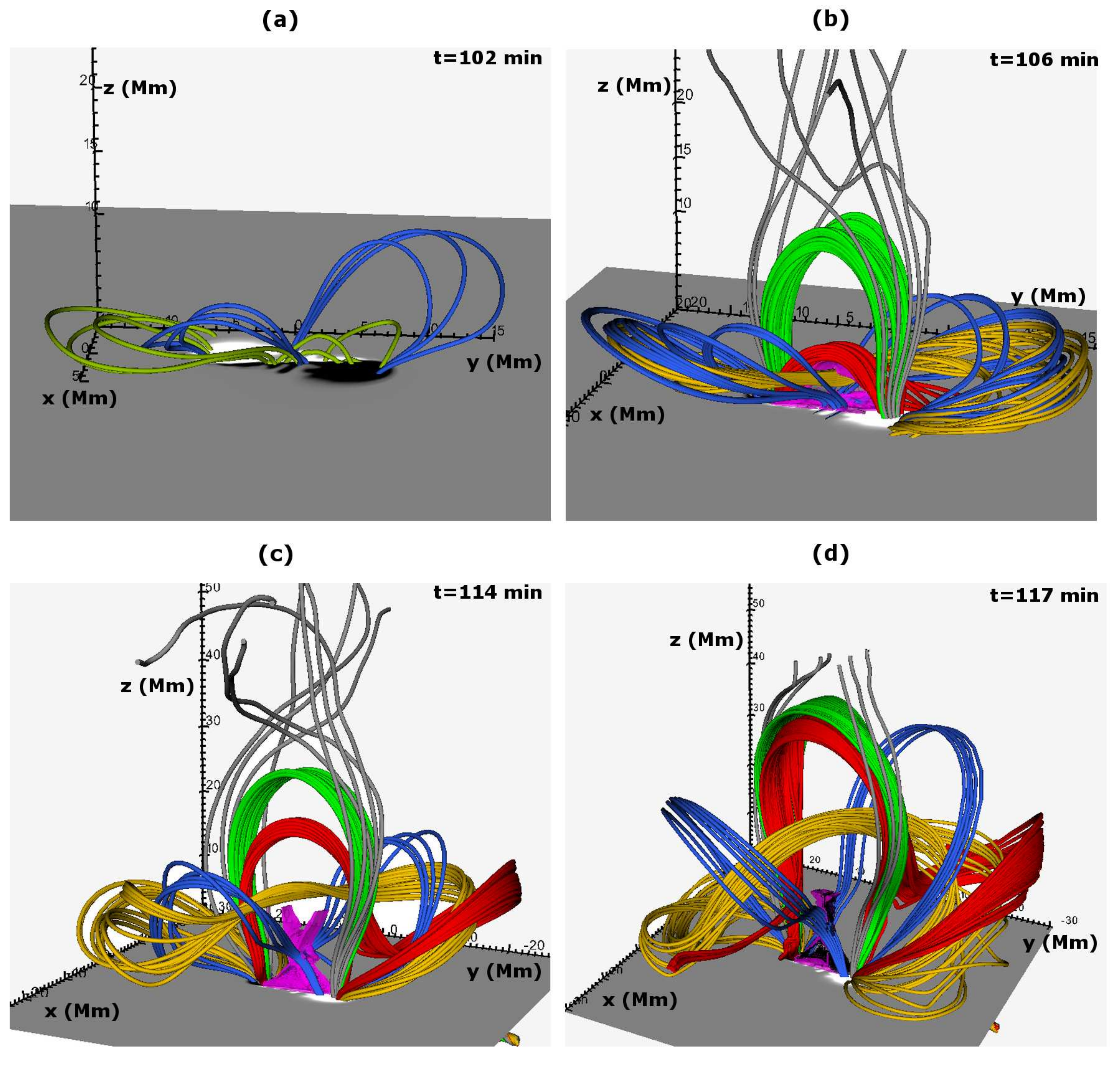}
\caption{ Field line morphology of the third eruption at $t$=102,106, 114, 117~min. \textbf{(a:)} J-like loops (blue) and sea-serpent fieldlines (dark green).
\textbf{(b,c,d):} similar to Fig.~\ref{fig:eruption1} and Fig.~\ref{fig:eruption2}.}
\label{fig:eruption3}
\end{figure*}
\begin{figure}
\centering
\includegraphics[width=0.90\columnwidth]{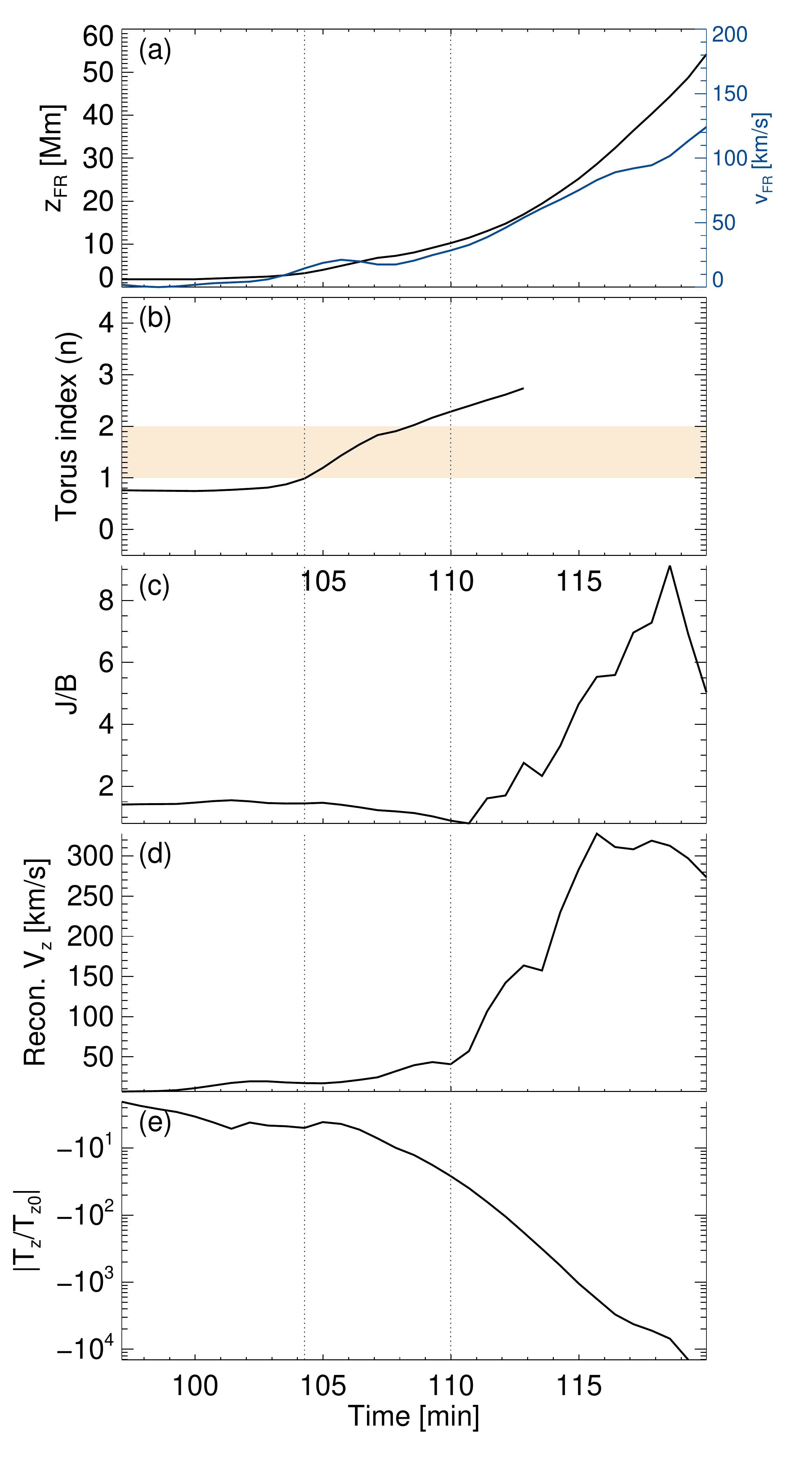}
\caption{Third eruption's key parameters with time.
\textbf{(a):} Height-time profile of FR center (black) and FR velocity (h-t derivative, blue) \textbf{(b):} Torus index measured along the height-time profile.  The highlighted region shows an estimated range of values for the occurrence of a torus instability. \textbf{(c):} Maximum $\left| J/B \right|$ along the CS. \textbf{(d):} Maximum reconnection outflow. \textbf{(e):} Ratio of mean tension ($T_z$) over its initial value ($T_{z_0}$). Vertical lines mark the times of the possible onset of the torus instability (first line) and the EJ-TC reconnection (second line).}
\label{fig:vars_eruption3}
\end{figure}

In the following, we focus on the dynamics of the second eruption. Fig.~\ref{fig:eruption2}a and Fig.~\ref{fig:eruption2}b are close-ups of the area underneath the first erupting FR at $t=74$~min and $t=79$~min respectively. In a similar manner to the formation of the first FR, the second FR (yellow fieldlines, Fig.~\ref{fig:eruption2}b) is formed due to reconnection between J-loops (blue fieldlines). The post-reconnection arcade (green and red fieldlines in Fig.~\ref{fig:eruption2}a, Fig.~\ref{fig:eruption2}b), which was formed after the first eruption (cyan lines, Fig.~\ref{fig:eruption1}f), overlies the yellow fieldlines and, thus, it acts as an envelope field for the second FR. Above and around this envelope field, there are fieldlines (grey) which belong to the first eruptive flux system but they have not exited the numerical domain yet. Hereafter, we refer to this field as the external, pre-existing field.

As the second post-emergence FR moves upwards, the envelope fieldlines are stretched vertically and their footpoints move towards the current sheet (pink isosurface). However, they do not reconnect in an EE-TC manner. Instead, the lower segments of the envelope fieldlines reconnect with the J-like loops. Hereafter, for simplicity, we refer to this as EJ-TC reconnection (i.e. Envelope-J Tether Cutting reconnection). This difference is due to the different orientation of the envelope fieldlines. As we have previously shown (green lines, Fig.~\ref{fig:4_plot}b top view), the envelope fiedlines in the second eruption do not have a strongly azimuthal nature. They are mainly oriented along the $y$-direction. Therefore, their lower segments come closer to the J-like loops and reconnect with them (e.g. bottom right red lines, Fig.\ref{fig:eruption2}c). 

To better illustrate the EJ-TC reconnection, in Fig.~\ref{fig:eruption2}d we show a close-up of this region. Here, the envelope fieldlines (green) reconnect with the J-like loops (blue) to form the hook-shaped fieldlines (red). 
Eventually, this process occurs on both foot points of the envelope fieldlines, forming new fieldlines such as the red ones in Fig.~\ref{fig:eruption2}e. 
Notice that these new reconnected fieldlines are winding around the footpoints of the rising FR and, therefore, they become part of the erupting field. 
In general, the EJ-TC reconnection removes flux from the envelope field and adds flux to the FR. Also, the downward tension of the envelope field decreases during EJ-TC reconnection. 
Before the FR exits the box (Fig.~\ref{fig:eruption2}f) most of the envelope field has been subject to EJ-TC reconnection. 
We should highlight that we don't find evidence of EE-TC reconnection during the second eruption.

EJ-TC and EE-TC reconnection produces fieldlines with a different shape. In the first eruption, the EE-TC reconnected fieldlines (red, Fig.~\ref{fig:eruption1}c-e) are ejected towards the FR center, adopting a ``W-shaped'' configuration. The concave-upward segments of the W-like fieldlines (arrows, Fig.~\ref{fig:eruption1}d) bring dense plasma from the low atmosphere into the central region of the FR. In the second eruption, the EJ-TC reconnected fieldlines have hook-like segments in their footpoints (arrows, Fig.~\ref{fig:eruption2}f). In this case, the tension of the reconnected fieldines ejects hot and dense plasma sideways (mainly along the y-direction) and not towards the center of the FR. Thus, due to the different way that the envelope fieldlines reconnect, the temperature and density distributions within the erupting field show profound differences, between the first and the following eruptions. This is discussed in more detail in Section 3.3.

We plot now the h-t profile and its derivative for the second FR (black and blue lines, Fig.~\ref{fig:vars_eruption2}a). 
To calculate the torus index, we consider the potential magnetic field $B_p$. 
As discussed earlier, the calculation of the potential field takes into account all the boundaries of the numerical domain. 
This means that the potential field solution will not approximate the envelope field everywhere. It will approximate the envelope field up to a height where the solution of the Laplace equation will be strongly influenced by the lower boundary (photosphere). 
Above that height, the potential solution will be influenced by the upper boundary and will describe the external field.
So, we examine the values of the potential field along height. We expect them not to change drastically in the region of the envelope field. 
We do find that the potential field solution does not describe the envelope field accurately above certain heights (different height for different snapshots). Below that heights, the potential field describes the envelope field well. This transition happens around $z\approx$15-20~Mm. Therefore, when we calculate the torus index, we do not take into account values of the torus index when the FR is located above $z$=15~Mm. 

According to the h-t profile, we find that the FR enters the exponential rise phase at $t=79.3$~min (first vertical line, Fig.~\ref{fig:vars_eruption2}b). The torus index at this time is $n=1.22$ and lies in the estimated range of values for the occurrence of a torus instability.
During this phase, the maximum $\left|J/B\right|$ does not increase dramatically (Fig.~\ref{fig:vars_eruption2}c). The current sheet becomes more elongated and the reconnection outflow becomes more enhanced after $t=81$~min (Fig.~\ref{fig:vars_eruption2}d).

When the EJ-TC reconnection starts, we find that the tension above the FR starts to decrease drastically (second vertical line, Fig.~\ref{fig:vars_eruption2}e). Also, after the initiation of the EJ-TC reconnection, the current density of the current sheet becomes more enhanced.

Due to the above, one possible scenario is that the torus instability is responsible for the onset of the exponential phase of the h-t profile, and the EJ-TC reconnection occurs {\it during} the rapid rise of the FR. Another possible scenario is that both processes are at work during the eruptive phase and it is the interplay between them, which leads to the fast eruption of the FR.

In terms of the energy, we have found that the kinetic energy of the second eruption is larger than that of the first eruption (red line, Fig.~\ref{fig:energy}). 
This difference is not necessarily associated with the different TC reconnection processes. For instance, the downward magnetic tension of the envelope field above the second FR is less. As a result, the upward motion of the FR is faster. Also, the photospheric unsigned magnetic flux increases between the two eruptions due to the continuous emergence. Thus, there is more available flux at the photosphere for the second eruption. Similarly, the magnetic energy in the corona (black line, Fig.~\ref{fig:energy}) increases between the two eruptions, indicating that more energy is available for the second eruption.


\subsubsection{Third and fourth eruption}

After the second FR exits the numerical domain, the overall fieldline morphology is similar to the first post-eruption phase. There is an external field, a post-reconnection arcade that acts as an envelope field and also the J-like loops.
At the photosphere-chromosphere, we also find sea serpent fieldlines (dark green lines, Fig.~\ref{fig:eruption3}a), similar to the previous work by \citet{Fan_2009, Archontis_etal2013}.
Most of these fieldlines originate from the partial emergence of the sub-photospheric field at different locations along the PIL. 
These fieldlines reconnect at many sites along the PIL during the early FR formation. Still, the major role in the FR formation is played by the reconnection of J-like loops (blue and yellow lines, Fig.~\ref{fig:eruption3}b)

In comparison to the second eruption, we find that the morphology of the external field is different. The second eruption (with a kinetic energy peak at $t$=87~min) happened right after the first eruption (with a kinetic energy peak at $t$=72~min). Thus, the external field that the second eruption had to push through was more horizontal (Fig.~\ref{fig:eruption2}c, gray lines are almost parallel to the photosphere). The third eruption, during which the kinetic energy takes its maximum value at $t$=119~min) happens after the second FR exits the numerical box. As a result, the external field is more vertical to the photosphere and, consequently, it has a very small downward tension (gray lines, Fig.~\ref{fig:eruption3}b).

EJ-TC reconnection occurs also during the third eruption (Fig.~\ref{fig:eruption3}c). However, we find that only some of the envelope fieldlines reconnect in both their footpoints (Fig.~\ref{fig:eruption3}c,d), before they exit the numerical domain. The implication of this difference will be discussed in Sec.~\ref{sec:temp_etc}.
We do not find evidence of EE-TC reconnection during the third eruption.

Regarding the torus instability, we should mention that at $t\approx100-104$~min, the FR is located very close to the photosphere, at heights $z\approx1.5-3$~Mm. We find that the value of $B_p$ (and hence $n$) at these heights depends on the choice of the lower boundary (i.e. the exact height of the photospheric layer, which is used to calculate the potential field). Thus, the value of the torus index for heights up to $z\approx3$~Mm are different. Above that height, all the solutions converge. We conjecture that the main reason for the change in the values of $B_p$ and $n$ is the build-up of a complex external field after each eruption.  

However, from the height-time profile (Fig.~\ref{fig:vars_eruption3}a), we find that for $t\simeq 104$~min the FR is located just above $z\approx3$~Mm, where the value of the torus index is well defined. Also, we find that $n\geq 1$ for $t > 104$~min (first vertical line, Fig.~\ref{fig:vars_eruption3}b). This is an indication (although not conclusive) that the torus instability might be associated with the onset of the eruption.       

Notice that during the time period $t\approx104-110$~min, there is no direct evidence that effective reconnection (e.g. EJ-TC reconnection) is responsible for the driving of the eruption. Fig.~\ref{fig:vars_eruption3}c, d, e show that the reconnection upflow underneath the flux rope undergoes only a small increase (due to reconnection between J-like fieldlines) and $J/B$ experiences a limited drop. The tension of the envelope fieldlines decreases mainly because of the 3D-expansion and not because of vigorous EJ-TC reconnection. Therefore, due to the above limitations, we cannot reach a definite conclusion about the exact contribution of reconnection at the onset of the eruption in this initial phase. 

In contrast, for $t > 110$~min, there is a clear correlation between the increase of the reconnection outflow and $J/B$ and the decrease of the tension. This is due to effective EJ-TC reconnection, which releases the tension of the envelope field and it boosts the acceleration of the erupting field. A preliminary comparison between the second and third eruptions show that the the maximum values of the current and reconnection outflow are similar, while the length of the CS and the extend of the jet are much smaller. The fourth eruption is very similar to the third eruption.

\subsection{Temperature, density, velocity and current}
\label{sec:temp_etc}
\begin{figure*}
\centering
\includegraphics[width=0.93\textwidth]{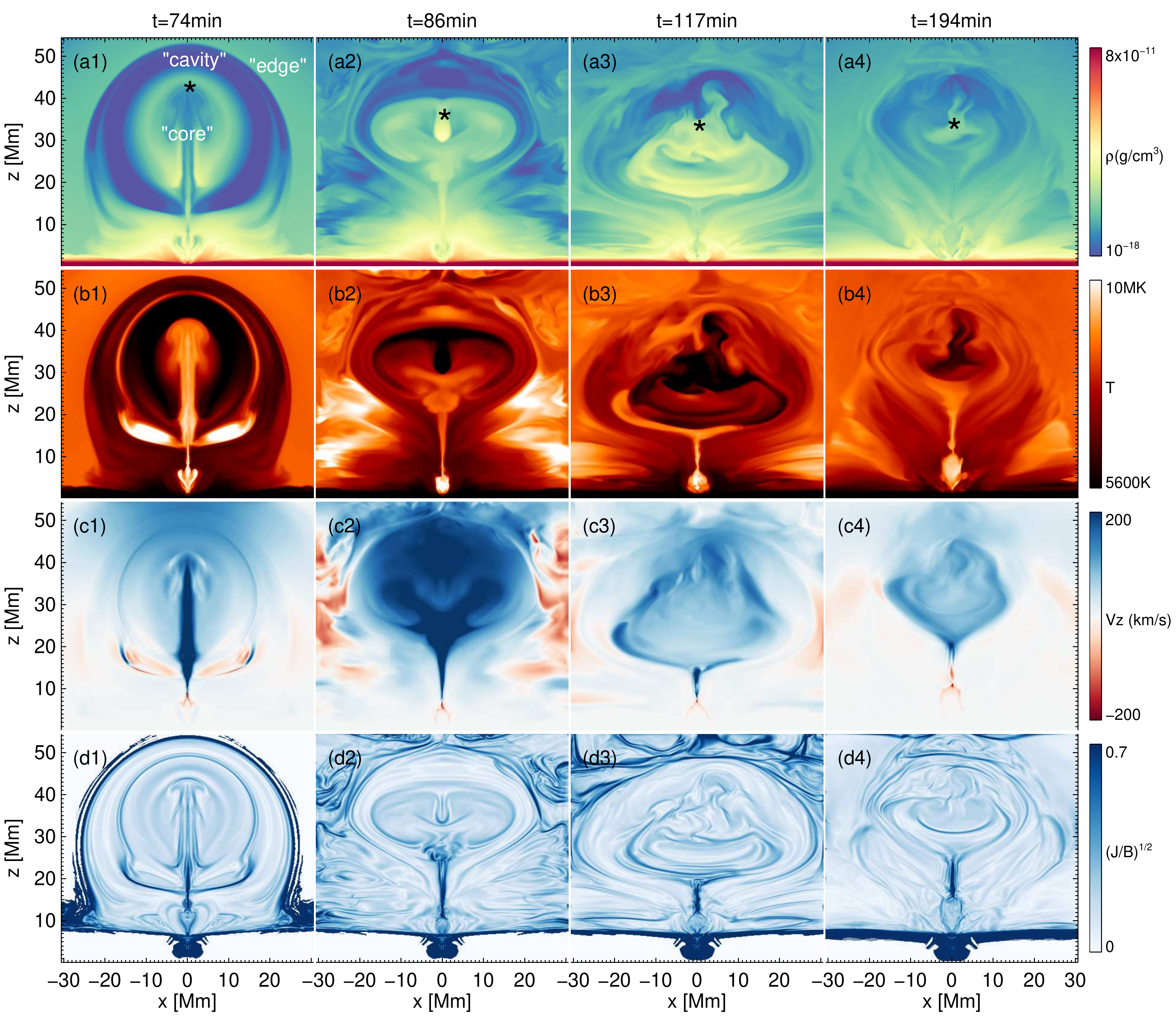}
\caption{Density (first row), temperature (second row), $V_z$ (third row) and $\sqrt{\left| J/B \right|}$ (fourth row) measured at the $xz$-midplane, for the first (first column), second (second column), third (third column) and fourth (fourth column) eruption. Asterisks mark the location of the FR center.}
\label{fig:temp_etc}
\end{figure*}

There are some remarkable similarities and differences between the four eruptions, as illustrated in 
Fig.~\ref{fig:temp_etc}. All panels in this figure are 2D-cuts, at the vertical $xz$-midplane, at times just before the erupting structures exit the numerical domain.

The density distribution (first row) shows that all eruptions adopt an overall bubble-like configuration, due to the expansion of the magnetic field as it rises into larger atmospheric heights. We notice that the erupting field consists of three main features, which are common in all eruptions. For simplicity, we mark them only in the first eruption (panel a1). These features are: (a) the inner-most part of the bubble, which is located at and around the center of the erupting field (marked by asterisk), filled with dense plasma - we refer to this part as the ``core'' of the eruption, (b) the low density area that immediately surrounds the ``core'' -  we refer to this as the ``cavity'' and it is the result of the cool adiabatic expansion of the rising magnetic field and (c) the ``front'' of the erupting structure, which is a thin layer of dense material that envelops the ``cavity'' and it demarcates the outskirts of the erupting field. To some extent, the shape of the eruptions in our simulations is reminiscent of the ``three-part'' structure of the observed small-scale prominence eruptions \citep[mini or micro CMEs e.g.][]{Innes_etal2010b,Raouafi_etal2010,Hong_etal2011} and/or CMEs   \citep[e.g.][]{Reeves_etal2015}. Because of this, hereafter, we refer to the simulated eruptions as CME-like eruptions.

Now, by looking at the temperature distribution (second row), we notice that there is a mixture of cold and hot plasma within the erupting field (in all cases, b1-b4). In fact, in the first eruption, there is a noticeable column of hot plasma, which extends vertically from $x=0$~Mm, $z=10$~Mm up to $z=40$~Mm. Thus, in this case the ``core'' of the erupting field appears to be hot, with a temperature of about 8~MK. On the contrary, the ``core'' of the following eruptions is cool (5,000-20,000~K) and dense, but is surrounded by hot (0.5-2~MK) plasma. In all cases, the origin of the hot plasma is the reconnection process occurring at the flare current sheet underneath the erupting field. 
The distribution of $\sqrt{\left| J/B \right|}$ is shown at the fourth row in Fig.~\ref{fig:temp_etc} (d1-d4). 
The flare current sheet is the vertical structure with high values of $\sqrt{\left| J/B \right|}$, and is located at around $x=0$~Mm and between $z=12$~Mm and $z=25$~Mm. The velocity distribution (panels c1-c4) shows that a bi-directional flow is emitted from the flare current sheet. This flow is a fast reconnection jet, which transfers the hot plasma upwards (into the erupting field) and downwards (to the flare arcade located below $z=10$~Mm). 

Thus, a marked difference between the first and the following eruptions is that, in the first eruption, the upward reconnection jet shoots the hot plasma vertically into the ``core'' of the erupting field, while in the following eruptions, the upward jet only reaches lower heights, arriving below the ``core''. 
In the latter cases, the jet is diverted sideways at heights below the center of the erupting FR, adopting a Y-shaped configuration (e.g. see c2-c4). In the first eruption, the EE-TC reconnection creates fieldlines which have a highly bended concave-upward shape (i.e. towards the central region of the erupting bubble, see red lines in Fig.~\ref{fig:eruption1}d). It is the strong (upward) tension of these fieldlines, that makes the hot plasma to be ejected at large heights and into the ``core'' of the field. In the following eruptions, the tension force that accelerates the hot jet upflow is weaker. This is because the reconnected fieldlines of the jet is the result of reconnection between Js (e.g. see blue lines in Fig.~\ref{fig:eruption2}e), which are not so vertically stretched as the envelope fieldlines during the EE-TC reconnection. Thus, the upward tension of the reconnected fieldlines at the flare current sheet is weaker. Therefore, the hot reconnection jet is not strong enough to reach large atmospheric heights and to heat the central region of the erupting field. When it reaches close to the heavy core of the erupting FR, it is diverted sideways (where the pressure is lower) and the embedded hot plasma runs along the reconnected fieldlines. 

In general, the temperature distribution within the overall volume of the erupting field correlates well with the distribution of $\sqrt{\left| J/B \right|}$, which implies that heating occurs mainly at sites with strong currents. As we mentioned above, one such area is the flare current sheet underneath the erupting FR. Another example is the heating that occurs at a thin current layer formed between the ``core'' and the ``cavity'' of the erupting bubble (e.g., see panel b1). This current layer is formed after the onset of the fast-rise phase of the erupting FR. The uppermost fieldlines of the erupting core are moving upwards with a higher speed than the fieldlines within the ``cavity'', which rise due to the expansion of the emerging field. Thus, at the interface between the two sets of fieldlines, the plasma is compressed and it is heated locally (up to 1~MK). This process occurs in all cases (panels, d1-d4), although it is more clearly visible in the first eruption (panel b1). The reconfiguration of the field after the first eruption leads to a more complex fieldline morphology, distribution of $\sqrt{\left| J/B \right|}$ and heating within the rising magnetized volume (panels b1-b4).         


\subsection{Geometrical extrapolation}
\label{sec:extrapolation}

\begin{figure*}
\centering
\includegraphics[width=0.98\textwidth]{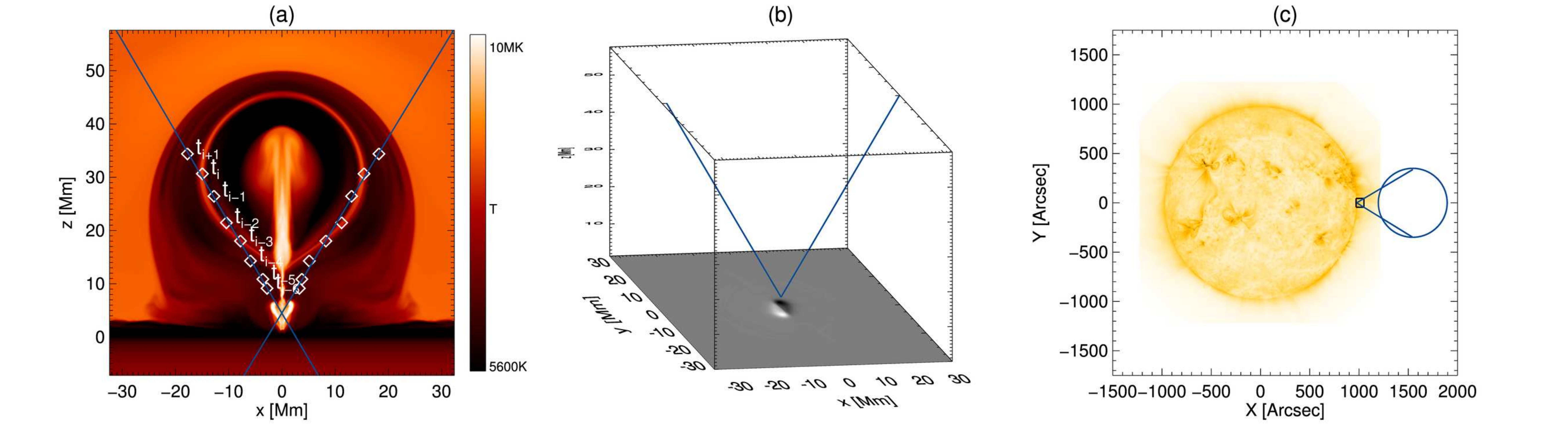}
\caption{ \textbf{(a):} Geometrical extrapolation based on the position of the flanks of the magnetic volume during its eruption. \textbf{(b):} Same as (a) but shown in the 3D volume of the numerical domain. \textbf{(c):} The extrapolated size of the erupting volume at 0.6~R$_\odot$ above the solar surface. The black box has the physical size of the simulation box.}
\label{fig:extrapolation}
\end{figure*}

Coronographic observations of CMEs show that they usually exhibit a constant angular width (i.e the flanks of the erupting structure move upward, along two approximately straight lines) \citep[e.g.][]{Moore_etal2007}.
Based on that, we perform a geometrical extrapolation of the size of the first eruption. 
For this, we find the location of the flanks of the structure at consecutive times and fit a straight line. Firstly, we mark the location of the flank of the erupting structure at a time $t_i$, when the flank is very distinguishable (diamond on the left flank, Fig.~\ref{fig:extrapolation}a). Next, we select the flank location prior to $t_i$ (marked with $t_{i-6},t_{i-5}$ etc.) and after (marked with $t_{i+1}$), and fit a straight line through these points (blue line). We then do the same for the other flank. The point where they intersect is approximately the height of the initiation. These extrapolated lines are also plotted in the 3D volume of our numerical box for better visualization (Fig.~\ref{fig:extrapolation}b). 

After we find these lines, we extrapolate them to 0.6~$R_\odot$. For size comparison, we plot them on the solar limb (blue lines, Fig.~\ref{fig:extrapolation}b). The box at the bottom of the extrapolations shows the size of our numerical box. It is clear that although the eruptions originate from a small-scale region, they grow in size, and it is not unlikely that they may evolve into considerably larger-scale events.
We should highlight that the above method is a first order approximation regarding the spatial evolution of the first eruption, assuming that the erupting field will continue to rise and expand even after it leaves the numerical domain.

The maximum value of the magnetic energy in the simulated eruptions is $1\times10^{28}$~erg and the kinetic energy varies in the range $3\times10^{26}-1.5\times10^{27}$~erg. Based on the size of our numerical box and the aforementioned values of energies, the eruptions in this simulation could describe the formation and ejection of small scale CME-like events. Most CMEs have typical values of kinetic energies around $10^{28}-10^{30}$~erg \citep{Vourlidas_etal2010}. 

\section{Summary and discussion}
\label{sec:conclusions}

In this work we studied the formation and triggering of recurrent eruptions in an emerging flux region using numerical simulations. The initial emergence of the sub-photospheric flux tube formed a bipolar region at the photosphere.  The combination of shearing motions and the rotation of the two opposite polarities formed J-like fieldlines, which reconnected to create a FR that eventually erupted ejectively towards the outer solar atmosphere.
In total, four successive eruptions occurred in the simulation. 
We found that the strength of the magnetic envelope field above the eruptive FRs dropped fast enough so that the FRs became torus unstable.

The initial slow-rise phase of the first FR started due to the torus instability. The rising FR pushed the envelope field upwards. The fieldlines of the envelope field reconnected in a tether-cutting manner and, as a result, the tension of the overlying field dropped in an exponential way. 
At that time, the FR entered the fast-rise phase. The fieldlines formed due to the reconnection of Js, turned about one time around the axis of the FR, while the fieldlines resulting from the tether-cutting of the envelope field turned about at least two times around the axis of the FR. The reconnected fieldlines that were released downwards, formed a post-reconnection arcade.

After the eruption of the first FR, reconnection of J-like fieldlines continued to occur and another FR was formed, which eventually erupted. This process of FR formation occurred two more times in a similar manner. In all cases, the post-reconnection arcade acted as a new ``envelope'' field for the next FR. We found that the envelope field was decaying fast enough to favor torus instability. The envelope fields between the second, third and fourth eruption differed mostly at the height where the FRs became torus unstable ($n\approx1-2$).
However, we should highlight that our calculation of the torus index is approximate because the envelope field evolves dynamically (e.g. it undergoes expansion). The derivation of the torus instability criteria based on previous analytical studies, took into account perturbations of a static configuration. Thus, a more accurate estimate of the torus index in our simulations, would be to let the envelope field to relax at each time step and then calculate $n$. This can only be done if the driver of the system could be stopped, letting the overall magnetic flux system to reach an equilibrium \citep[e.g.][]{Zuccarello_etal2015}.
However, in our dynamical simulations, there is a certain amount of available magnetic flux, which can emerge to the photosphere and above. The driver of the evolution of the system (i.e. magnetic flux emergence) cannot be stopped before the available magnetic flux is exhausted. Therefore, on this basis, we study the {\it continuous} evolution of the system. 
Still, in our experiments, the magnitude of the current inside the envelope field is at least ten times lower than the one in the FR core, so we expect that the envelope field is not far away from the potential state.

The removal of the downward tension of the envelope field is important for the erupting FRs. In the first eruption, the removal of the envelope tension occurred through the reconnection of the envelope field with other envelope fieldlines (EE-TC reconnection). In the other three eruptions, the envelope field reconnected with J-like fieldlines (EJ-TC reconnection). The differences between EE-TC reconnection and the EJ-TC reconnection were found to be significant for the density and temperature distribution within the erupting structure. After the EE-TC reconnection, the reconnected fieldlines underneath the erupting FR adopted a W-like shape, with two upward concave regions (red lines Fig.~\ref{fig:eruption1}d, see arrows). After the EJ-TC reconnection, the lower segments of the reconnected fieldlines adopted a hook-like shape (red lines, Fig.~\ref{fig:eruption2}f, see arrows).

In the case of EE-TC reconnection, the upward tension of the reconnected fieldlines (as illustrated by the upward-stretched segments in the middle of the W-shaped fieldlines) pushed hot plasma from the flare current sheet into the erupting field via a hot and fast collimated jet. Due to this process, the temperature of the central region of the erupting FR changed during the eruption, from low to high values (b1, Fig.~\ref{fig:temp_etc}). 

In the case of EJ-TC reconnection, the plasma transfer from the flare current sheet to the erupting field was mainly driven by the reconnection of Js, and therefore the resulting reconnection jet was not as collimated as on the EE-TC reconnection. In the second eruption, this post-reconnection hot jet collided with the FR and became diverted into two side jets (a2 and c2, Fig.~\ref{fig:temp_etc}). In the third and fourth eruption, the jets were not fast enough to enter the region of the erupting core of the field (a3 and c3, a4 and d4, Fig.~\ref{fig:temp_etc}).

Thus, the study of the temperature distribution revealed that due to EE-TC reconnection, the erupting field develops a ``3-part'' structure consisted of a hot front ``edge'', a cold ``cavity'', and a hot and dense ``core''. In the following eruptions, the temperature of the plasma within the central region of the FRs remained low. Therefore, we suggest that the observations of erupting FRs, which are heated e.g. from $10^{3}$~K to$10^{6}$~K, \textit{during} their eruptive phase, might indicate that EE-TC reconnection is at work. We should mention that heat conduction is not included in our simulation. 
Therefore, the exact value of the temperature within the erupting field may change if heat conduction were to be included in the numerical experiment.

Overall, we report that the physical mechanism behind the formation of recurrent ejective eruptions in our flux emergence simulation is a combination of torus unstable FRs and the onset of tether-cutting of the overlying field through a flare current sheet. 
Both the EE-TC reconnection and the EJ-TC reconnection were found to remove the downward tension of the overlying field and thus assisting the eruptions. In the first eruption, it is likely that torus instability occurs first, and the rapid exponential rise phase of the erupting FR comes after the EE-TC reconnection. For the other eruptions, where the structure of the magnetic field above the FR has a more intricate morphology, it is difficult to conclude which process is responsible for the onset of the various phases of the eruptions.

Comparing our results with previous studies, the formation of all the FRs in our simulation is due to the reconnection of sheared J-like fieldlines, in a similar manner to earlier simulations \citep[e.g. ][]{Aulanier_etal2010,Archontis_etal2012,Leake_etal2013,Leake_etal2014}. 

It is also interesting to note that the velocity and current profile of our first eruption (Fig.~\ref{fig:temp_etc}c1, d1) are very similar morphologically to the ones produced from the flare reconnection in the breakout simulation of \citet{Karpen_etal2012}, who used a (different) 2.5D adaptive grid code. 
Such similarities indicate that the resulting morphologies might be generic and indicative of the EE-TC reconnection.

\citet{Moreno-Insertis_etal2013} performed a flux emergence simulation of a highly twisted flux tube into a magnetized atmosphere and found recurrent eruptions. In comparison to our simulation, the sub-photospheric flux tube in the work of \citet{Moreno-Insertis_etal2013} had higher magnetic field strength ($B_{0}=3.8$~kG), higher length of the buoyant part of the flux tube ($\lambda$=20 in comparison to our $\lambda$=5) and was located closer to the photosphere ($z=-1.7$~Mm). In their work, their first FR is formed, similarly to our simulation, by the reconnection of sheared-arcade fieldlines. The higher $\lambda$ leads to the formation of a more elongated emerging FR and a longer sigmoid. The eruption mechanism, though, is very different. It involves reconnection between the sheared-arcade fieldlines and the open fieldlines of the ambient field. Also, it involves reconnection of the sheared-arcade with a magnetic system produced from the reconnection of the ambient field with the initial emerging envelope field. Their second and third eruption are off-centered eruptions of segments of the initial flux tube, that eventually become confined by the overlying field. In our case, the flux tube axis emerges only up to 2-3 pressure scale heights above the photosphere ($z$=0) and the erupting FRs are all formed due to reconnection of J-loops.

\citet{Murphy_etal2011} discussed possible heating mechanisms for the dynamic heating of CMEs, one of which is heating from the CME flare current sheet. Taking into account the results of previous studies \citep[e.g. ][]{Lin_etal2004}, they reported that the reconnection hot upward jets from the flare current sheet could reach the cool central region of the erupting FR and heat it. In fact, this leads to some mixing of hot and cool plasma within the central erupting volume.

From the two different tether-cutting reconnections found in our simulation, only the EE-TC reconnection allows effective transfer of hot plasma from the flare current sheet into the FR central region, by the reconnection outflow. 
This might account for a process similar to the afore-mentioned mixing of hot and cold plasma, as suggested by \citet{Lin_etal2004}. 
On the other hand, during EJ-TC reconnection, hot plasma is mainly found at the periphery of the central region of the FR.

The physical size of our simulated emerging flux region was 23.4~Mm, and the size of the FRs was up to 64.8~Mm (the length of the $y$-axis). The height of our numerical box was 57.6~Mm. The kinetic energies of the eruptions were $3\times10^{26}-1.5\times10^{27}$~erg and the magnetic energies around $1\times10^{28}$~erg. 
These values suggest that our numerical experiment describes an emerging flux region, which hosts relatively low energy eruptions in comparison to CMEs. Based on the sizes and the energetics, these eruptions can describe the formation and eruption of small scale eruptive events. For instance, such an eruption in terms of physical size and not magnetic configuration, was reported by \citet{Raouafi_etal2010,Reeves_etal2015}. Still, the results on the plasma transfer for the different flare reconnections (EE-TC reconnection and EJ-TC reconnection) should be scale invariant.

Having reproduced a CME-like configuration (a1 and b1, Fig.~\ref{fig:eruption1}) we extrapolated the expansion of the flanks of the erupting ``bubble'' and estimated its size in  0.6~R$_\odot$. We found that these eruptions have the potential to become comparable to small-sized CMEs (Fig.~\ref{fig:extrapolation}c), but with one order of magnitude lower kinetic energy. 
We aim to study the parameters that would increase the energies of the produced eruptions. For this, in our next paper, we will present the results of a parametric study on the magnetic field strength of the subphotospheric flux tube. Our aim is to study the differences in energetics, physical size and recurrence of the eruptions.

\acknowledgments
The Authors would like to thank the Referee for the constructive comments.
This project has received funding from the Science and Technology Facilities Council (UK) through the consolidated grant ST/N000609/1.
This research has been co-financed by the European Union (European Social Fund - ESF) and Greek national funds through the Operational Program ``Education and Lifelong Learning'' of the National Strategic Reference Framework (NSRF) - Research Funding Program: Thales Investing in knowledge society through the European Social Fund.
The authors acknowledge support by the Royal Society.
This work was supported by computational time granted from the Greek Research \& Technology Network (GRNET) in the National HPC facility - ARIS. 
This work used the DIRAC 1, UKMHD Consortium machine
at the University of St Andrews and the DiRAC Data Centric system at
Durham University, operated by the Institute for Computational Cosmology on
behalf of the STFC DiRAC HPC Facility (www.dirac.ac.uk). This equipment
was funded by BIS National E-infrastructure capital grant ST/K00042X/1, STFC
capital grant ST/H008519/1, and STFC DiRAC Operations grant ST/K003267/1
and Durham University. DiRAC is part of the National E-Infrastructure.

\bibliographystyle{apj}
\bibliography{bibliography}

\clearpage

\end{document}